\documentclass{aa}  

\usepackage{graphicx}
\usepackage{txfonts}

\usepackage{color}
\usepackage{hyperref}

\hypersetup{
    colorlinks=true,
    linkcolor=blue,
    urlcolor=blue,
    citecolor = blue,
}

\usepackage{tablefootnote}

\begin{document} 

   \title{The AMBRE Project: Solar neighbourhood chemodynamical constraints on Galactic disc evolution}
   
   \author{P. Santos-Peral
          \inst{1}
          \and
          A. Recio-Blanco\inst{1}
          \and
          G. Kordopatis\inst{1}
          \and
          E. Fernández-Alvar\inst{1}
          \and
          P. de Laverny\inst{1}
          }

   \institute{\inst{1} Université Côte d'Azur, Observatoire de la Côte d'Azur, CNRS, Laboratoire Lagrange, Bd de l'Observatoire, CS 34229, 06304 Nice cedex 4, France\\
              \email{psantos@oca.eu}
             }

   \date{Received December 16 2020 / Accepted June 16 2021}

 
  \abstract
   {The abundance of $\alpha$-elements relative to iron ([$\alpha$/Fe]) is an important fossil signature in Galactic archaeology for tracing the chemical evolution of disc stellar populations. High-precision chemical abundances, together with accurate stellar ages, distances, and dynamical data, are crucial to infer the Milky Way formation history.}
   {The aim of this paper is to analyse the chemodynamical properties of the Galactic disc using precise magnesium abundance estimates for solar neighbourhood stars with accurate \emph{Gaia} astrometric measurements.}
   {We estimated ages and dynamical properties for 366 main sequence turn-off (MSTO) stars from the AMBRE Project using PARSEC isochrones together with astrometric and photometric values from \emph{Gaia} DR2. We use precise global metallicities [M/H] and [Mg/Fe] abundances from a previous study in order to estimate gradients and temporal chemodynamic relations for these stars.} 
   {We find a radial gradient of -0.099~$\pm$~0.031~dex~kpc$^{-1}$ for [M/H] and +0.023~$\pm$~0.009~dex~kpc$^{-1}$ for the [Mg/Fe] abundance.  The steeper [Mg/Fe] gradient than that found in the literature is a result of the improvement of the AMBRE [Mg/Fe] estimates in the metal-rich regime. In addition, we find a significant spread of stellar age at any given [Mg/Fe] value, and observe a clear correlated dispersion of the [Mg/Fe] abundance with metallicity at a given age. While for [M/H]~$\leq$~-0.2, a clear 
   age--[Mg/Fe] trend is observed, more metal-rich stars display ages from 3 up to 12~Gyr, describing an almost flat trend in the 
   [Mg/Fe]--age relation.  Moreover, we report the presence of radially migrated and/or churned stars for a wide range of stellar ages, although we note the large uncertainties of the amplitude of the inferred change in orbital guiding radii. Finally, we observe the appearance of a second chemical sequence in the outer disc, 10-12 Gyr ago, populating the metal-poor, low-[Mg/Fe] tail. These stars are more metal-poor than the coexisting stellar population in the inner parts of the disc, and show lower [Mg/Fe] abundances than prior disc stars of the same metallicity, leading to a chemical discontinuity. Our data favour the rapid formation of an early disc that settled in the inner regions, followed by the accretion of external metal-poor gas ---probably related to a major accretion event such as the \emph{Gaia}-Enceladus/Sausage one--- that may have triggered the formation of the thin disc population and steepened the abundance gradient in the early disc.}
   {}


   \maketitle
%

\section{Introduction}

A thorough understanding of the formation and evolution of the Milky Way demands precise chemical abundances and stellar ages. The stellar upper atmospheres of non-evolved stars provide fossil evidence of the available metals of the interstellar medium (ISM) at the time of formation \citep{freeman2002}. The present-day observed chemical signatures in the Galactic stellar populations, together with ages, distances, and dynamical data, allow us to infer the different stages of the history of the Galaxy. \par

The formation of the Galactic disc is still not well understood and in particular the origin and existence of a thin--thick disc bimodality is a matter of debate. Following the initial thick disc identification from stellar density distributions \citep{yoshii1982, gilmorereid1983} and the first attempts to  kinematically classify the stars as part of either the thin or the thick disc \citep{bensby2003, bensby2005, reddy2006}, the two Galactic disc populations in the solar neighbourhood are often distinguished based on their abundances of $\alpha$-elements  (e.g. Mg, Si, Ti) relative to iron 
\citep[e.g.][]{vardan2012,alejandra2014,bensby2014,georges2015b,georges2017, wojno2016, fuhrmann2017,hayden2017,minchev2018}. The [$\alpha$/Fe] versus metallicity [M/H] plane provides valuable clues as to the disc stellar population evolution. The thick disc is often reported to be [$\alpha$/Fe]-enhanced relative to the thin disc, suggesting distinct chemical evolution histories. 
In particular, magnesium is often used as an $\alpha$-elements tracer \citep[e.g.][]{fuhrmann1998, sarunas2014,bergemann2014,carrera2019} because there is a high number of measurable spectral lines in optical spectra, and it also clearly separates the chemical sequences of the discs. Additionally, observational studies have shown that the thick disc stellar population could have been formed on a short timescale before the epoch of the thin disc formation \citep{haywood2013,georges2015b,fuhrmann2017,silvaaguirre2018,delgadomena2019,ciuca2020, katz2021}. This scenario is supported by two-infall chemical evolution models \citep{grisoni2017, spitoni2019, palla2020}.   \par

Moreover, the observed stellar radial abundance distribution and the age--abundance relations in the Galactic disc are interesting signatures for studying the chemical enrichment history. They contain information about the star formation efficiency at different Galactocentric distances and on different timescales \citep{magrini2009,minchev2014,anders2014}, the radial migration of stars \citep{sellwood2002,schonrich2009,minchev2018}, and the infall of gas \citep{oort1970,schonrich2017,grisoni2017}. An inside-out formation scenario \citep{matteucci1989} combined with a higher star formation efficiency in the inner regions of the Galaxy seems to reproduce the present-day local abundance gradients \citep{grisoni2018}, where metallicity decreases with Galactocentric radius. In addition, a tight correlation between [Mg/Fe]-enhancement and age has been found for the old thick disc population \citep[e.g.][]{haywood2013, bensby2014, hayden2017,delgadomena2019, nissen2017, nissen2020}, but without global consensus \citep[e.g.][]{silvaaguirre2018}. However, the stellar thin disc component shows a larger dispersion, along with a significant scatter in the age--metallicity relation (AMR). This signature has been found to be a sign of the superposition of different stellar populations in the thin disc, with different enrichment histories and birth radii \citep{carlberg1985, friel1995, nordstrom2004, schonrich2009, wojno2016}. \par

Furthermore, the radial migration of stars in the Galaxy, through churning and blurring, is supported by theory \citep{sellwood2002,schonrich2009,diMatteo2013} and backed up by observational evidence provided by different spectroscopic stellar surveys such as RAVE \citep{georges2015a}, HARPS \citep{hayden2017,minchev2018}, and APOGEE \citep{georges2017,feltzing2020}. The presence of stars with  supersolar metallicity ([M/H] $\geq$ + 0.1~dex) and circular orbits at solar galactocentric distances has been interpreted as clear evidence of stellar migration from inner birth regions in the Galaxy. In addition, migrated stars are expected 
to cause a flattening of the radial abundance gradients with time \citep{prantzos1999, hou2000, roskar2008, schonrich2009, pilkington2012, minchev2018, magrini2017,vincenzo2020}. The temporal and spatial dimensions are therefore crucial for interpreting and clarifying the present-day chemodynamical relations in order to disentangle the formation and evolution of the Galactic disc. In this paper, we refer to radial migration as the churning process, or simply `churning'. \par 

In a previous analysis \citep{SantosPeral2020}, we showed a significant improvement in the precision of [Mg/Fe] abundance estimates by carrying out an optimisation of the spectral normalisation procedure, in particular for the metal-rich population ([M/H]~>~0). The followed methodology made it possible to highlight a decreasing trend in the [Mg/Fe] abundance even at supersolar metallicites, partly solving the apparent discrepancies between the observed flat trend in the metal-rich disc \citep{vardan2012, hayden2015, hayden2017, sarunas2017, buder2019} and the steeper slope predicted by chemical-evolution models \citep{chiappini1997, romano2010, spitoni2020, palla2020}. In this paper, we use these new [Mg/Fe] abundance measurements in order to study their impact on the reported chemodynamical features (radial chemical abundance gradients, role of churning, age--abundance relations), and therefore on the interpretation of the Galactic disc evolution. We used the high-spectral-resolution AMBRE:HARPS dataset \citep{DePascale2014}, and analysed 366 main sequence turn-off (MSTO) stars in the local solar neighbourhood (d~<~300~pc from the Sun), for which we estimated ages and kinematical and dynamical parameters using the accurate astrometric measurements of the \emph{Gaia} space mission. \par 

The paper is organised as follows. In Sect. \ref{data}, we introduce the observational data sample used for the analysis. In Sect. \ref{results_1}, we show the radial chemical abundance gradients with Galactocentric radius and explore the effects of churning in the thin disc sample. We present our results on [Mg/Fe] and metallicity as a function of stellar age and orbital properties in Sect. \ref{results_2}. In Sect. \ref{discussion}, we discuss the proposed scenario for the formation and evolution of the Galactic disc. We conclude with a summary in Sect.~\ref{summary}.

\section{Data} \label{data}

\subsection{The AMBRE:HARPS sample} \label{harps} 

The AMBRE Project, described in \citet{patrick2013}, is a collaboration between the Observatoire de la Côte d’Azur (OCA) and the European Southern Observatory (ESO) to automatically and homogeneously parametrise archived stellar spectra from ESO spectrographs: FEROS, HARPS, and UVES. The stellar atmospheric parameters (T$_{\rm eff}$, log(g), [M/H], [$\alpha$/Fe]) were derived by the multi-linear regression algorithm MATISSE \citep[MATrix Inversion for Spectrum SynthEsis,][]{alejandra2006}, using the AMBRE grid of synthetic spectra \citep{patrick2012}. Additionally, the AMBRE Project estimates the radial velocity (v$_{\rm rad}$) by a cross-correlation function between the observed spectra and the used synthetic templates.  \par

For the present paper, we only considered a subsample of the AMBRE:HARPS spectral dataset\footnote{The AMBRE analysis of the HARPS spectra comprises the observations collected from October 2003 to October 2010 with the HARPS spectrograph at the 3.6m telescope at the La Silla Paranal Observatory, ESO (Chile).} \citep[R $\sim$ 115000, described in][]{DePascale2014} that corresponds to 494 MSTO stars in the solar neighbourhood, selected and used in \citet{hayden2017}. 
These latter authors  made the sample selection by requiring M$_{J}$~<~3.75 and 3.6~<~log~g~<~4.4. The external uncertainties (estimated by comparison with external catalogues) on T$_{\rm eff}$ , log(g), [M/H], [$\alpha$/Fe], and v$_{\rm rad}$ are 93K, 0.26~cm~s$^{-2}$, 0.08~dex, 0.04~dex, and 1~km~s$^{-1}$, respectively. Relative errors from spectra to spectra are much lower. As mentioned above, the stellar [Mg/Fe] abundances were derived following the methodology described in \citet{SantosPeral2020}, where the spectral normalisation procedure was optimised for the different stellar types and each particular Mg line separately. These abundances present an overall internal error of around 0.02 dex and an average external uncertainty of 0.01~dex with respect to four identified Gaia-benchmark stars (18 Sco, HD 22879, Sun, and $\tau$ Cet) from~\citet{jofre2015}. \par

\subsection{Gaia DR2: photometry, astrometry, and distances} \label{crossmatch}

We used astrometric \citep{lindegren2018} and photometric data \citep[][full passband information for BP and RP]{evans2018} from the \emph{Gaia} DR2 catalogue \citep{gaia2018}, along with distances estimated by \citet{bailer-jones2018} from \emph{Gaia} DR2 parallaxes using a Bayesian approach. We point out that, as the analysed sample is within a radius of 300 pc around the Sun, the parallax uncertainties ($\sigma_\varpi$/$\varpi$ < 3\% for our stars) and the choice of prior have little impact on the distance results. For the same reasons, new astrometric information from \emph{Gaia} EDR3 will not affect our conclusions.


\begin{figure*}
\centering
\hspace*{-1.5cm}
\includegraphics[height = 55mm, width=0.36\textwidth]{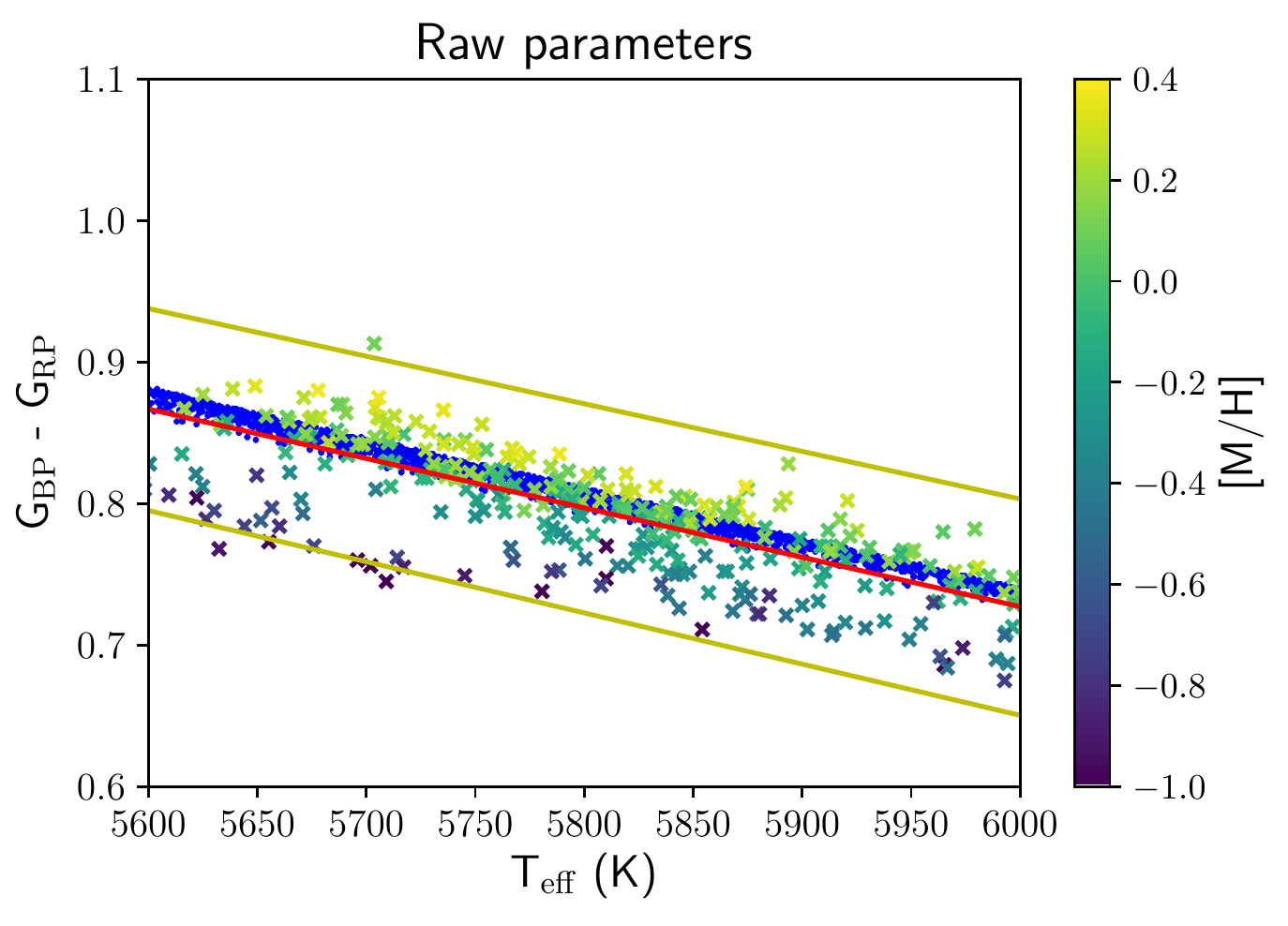}
\hspace{-0.2cm}
\includegraphics[height = 52mm, width=0.33\textwidth]{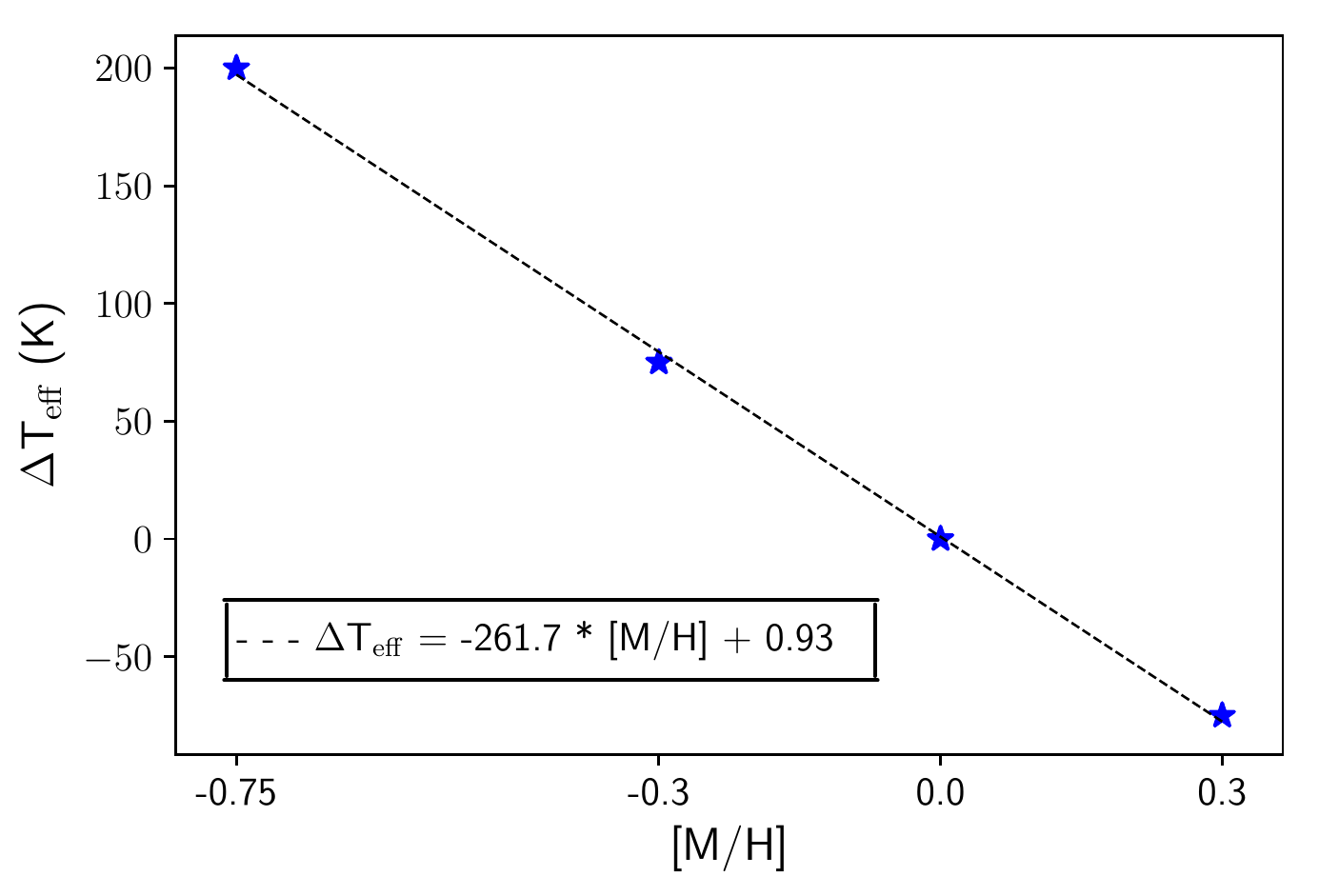}
\includegraphics[height = 55mm, width=0.36\textwidth]{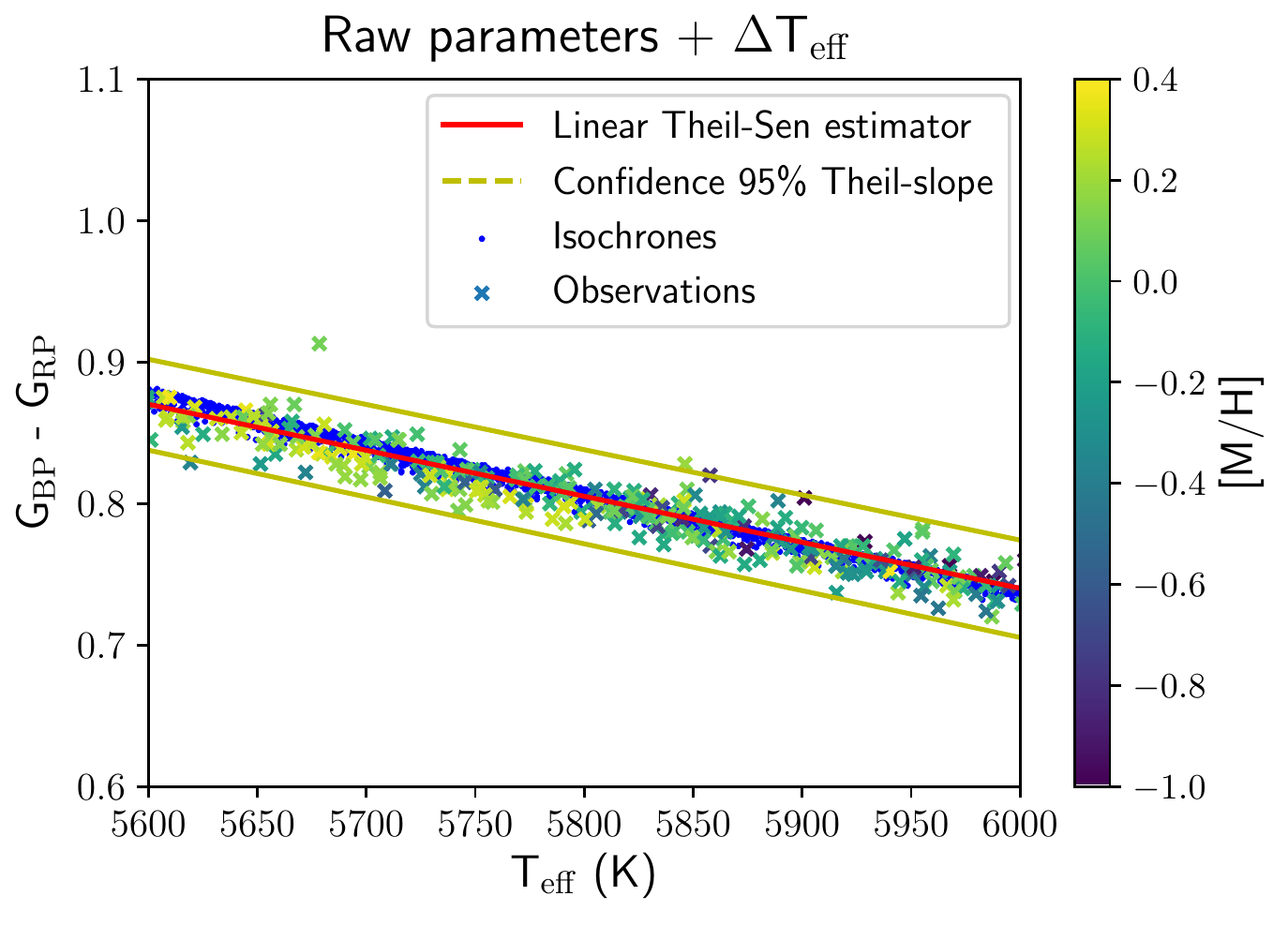}
\hspace*{-1.5cm}
\caption{Colour--temperature relation of the MSTO stars in the working sample (colour-coded according to metallicity) and the isochrone models (blue points on a straight line), for the original derived AMBRE effective temperatures (left panel) and after the offset correction in T$_{\rm eff}$ (right panel). The linear Theil-Sen estimator method (red line) was applied to the stellar parameter values. The green lines delimiter the lower and upper bound of the 95\% confidence interval of the Theil-Sen linear regression method. The middle panel shows the observed linear T$_{\rm eff}$--offset ($\Delta$T$_{\rm eff}$ = T$_{\rm eff,~MODEL}$ - T$_{\rm eff,~OBSERVATION}$) for different metallicity values.} 
\label{Fig:biasTeff}
\end{figure*}

We performed a cross-match of the whole AMBRE:HARPS database with the \emph{Gaia} DR2 catalogue through the CDS interface, looking for a match in a radius of 5~arcsec. This allowed us to assign a \textit{Gaia}~ID to each spectrum, identifying the different spectra of the same star. In order to verify the goodness of the cross-match, we compared the spectroscopically derived T$_{\rm eff}$ and radial velocities with those provided by \emph{Gaia} DR2. As the number of sources with T$_{\rm eff}$  determinations in \emph{Gaia} DR2 is not very large for the AMBRE:HARPS sample, we also estimated T$_{\rm eff}$ values from 2MASS \citep{2MASS} and APASS \citep{APASS} photometry, following \citet{gonzalez2009} (after performing the subsequent cross-match in the same way as we did with \emph{Gaia}), in order to compare with the spectroscopic ones. As an additional check, we also estimated the corresponding 2MASS and APASS photometric magnitudes from \emph{Gaia} photometry. Thus, we verified the correct identification of stars in the cross-match by comparing the T$_{\rm eff}$, the photometric magnitudes, and the radial velocities for each spectrum separately. We find that the median differences among AMBRE:HARPS, 2MASS, APASS, and Gaia data are lower than 150 K in T$_{\rm eff}$, and around 0.1 for the photometric magnitudes J, H, and K. As far as the radial velocities are concerned, we were only able to compare AMBRE:HARPS and \emph{Gaia} measurements, and find a median difference of about 0.02~km~s$^{-1}$. All these checks confirm that the AMBRE/\emph{Gaia} cross-match can be used with confidence. \par

Finally, we selected a cleaner spectra sample by excluding those spectra whose atmospheric parameters (T$_{\rm eff}$, log(g), [M/H], and [$\alpha$/Fe]) differ by more than two sigma from the mean value of the star. In addition, those stars with more than five observed spectra ($\geq$ 5 repeats) that present $\sigma_{\rm vrad}$ > 5 km~s$^{-1}$ were ruled out as binary system candidates. The remaining stars of the sample in the analysis present $\sigma_{\rm vrad}$ < 1 km~s$^{-1}$.

\subsection{Ages} \label{ages}

\begin{figure*}
\centering
\includegraphics[height=90mm, width=\textwidth]{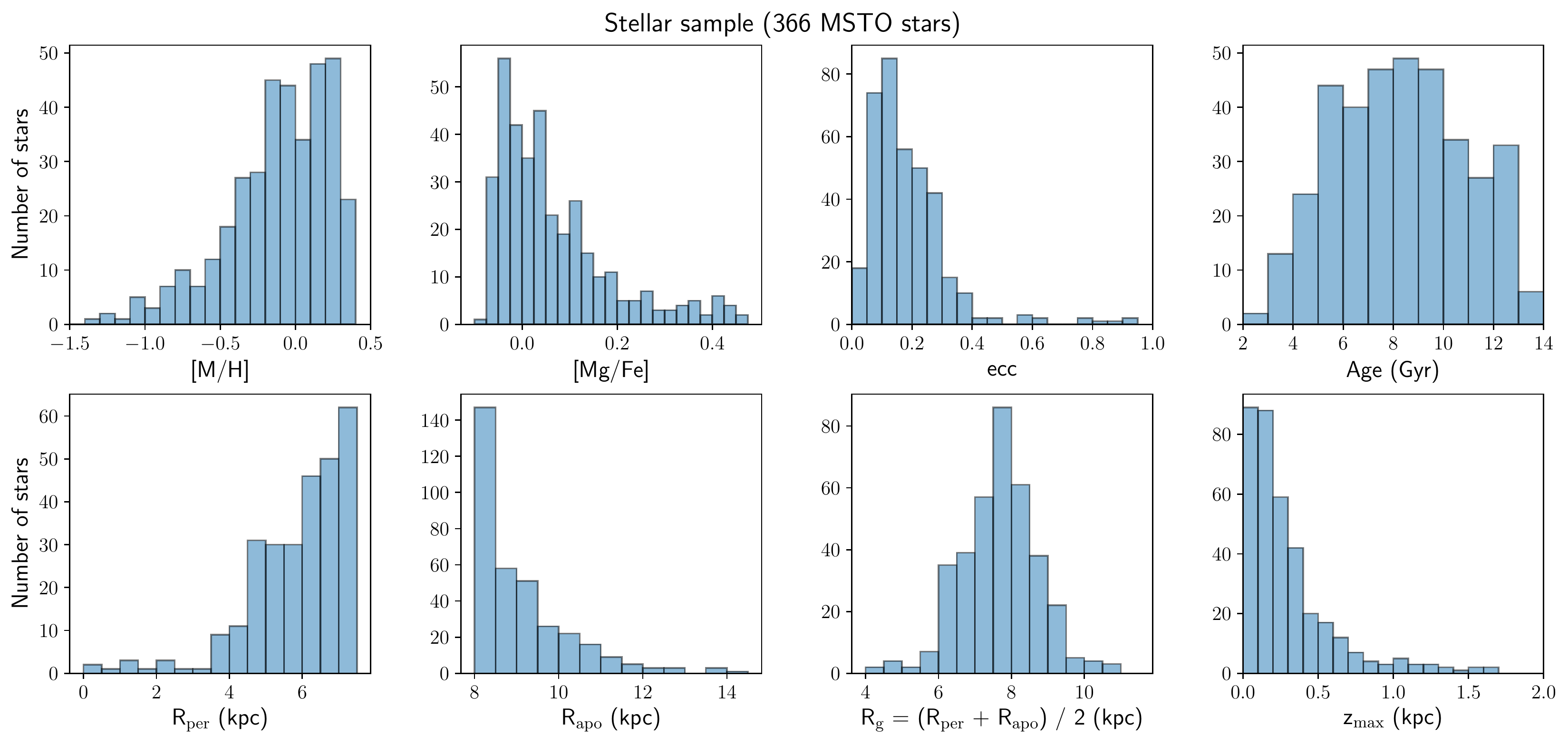}
\caption{Distribution of the main properties ([M/H], [Mg/Fe], eccentricity, age, R$_{\rm per}$, R$_{\rm apo}$, R$_{\rm g}$, z$_{\rm max}$) of the selected MSTO stars.}
\label{Fig:hist_ALL}
\end{figure*}

We restricted our sample to MSTO stars in order to estimate reliable ages using the isochrone-fitting method described in \citet{kordopatisAGES2016}. For each individual star, we computed the age probability distribution function (PDF) by projecting the stellar parameters (T$_{\rm eff}$, logg, [M/H]), the BP-RP colour, and the G absolute magnitude on the PARSEC isochrones \citep{bressan2012}, linearly scaled in age in steps of 0.1 Gyr from 0 to 15 Gyr, with the \citet{evans2018} colour transformation. Every of the parameters and magnitudes were weighted by their respective error bars, and each isochrone point was weighted by mass following \citet{zwitter2010}. \par

We first checked that the relation between the effective temperature (derived spectroscopically) and the colour (from \emph{Gaia}~DR2) was on the same scale as the one used by the isochrone models. As shown in the left panel of Fig. \ref{Fig:biasTeff}, a metallicity-dependent disagreement was found. The observed offset was evaluated for four different metallicity values\footnote{[M/H] = [ - 0.75 , - 0.3 , 0.0 , + 0.3]}, showing a linear dependence on the metallicity (middle panel):  

\vspace{-0.45cm}

\begin{equation} 
\centering
\rm \Delta T_{eff} \ (K) = \ (-261.7 \pm 5.4) \ * \ [M/H] \ (dex) \ \ + \ \ (0.93 \pm 2.32)
.\end{equation}

Figure~\ref{Fig:biasTeff} shows, for all the metallicities, the colour--temperature relation before (left panel) and after applying the correction in the AMBRE effective temperatures (right panel) in order to put them on the same temperature scale as the PARSEC isochrones. We determined the dispersion in the colour BP--RP differences between the observations and the models, finding an improvement from $\sigma_{\rm \Delta(BP-RP)}$ $\approx$ 0.028 to $\sigma_{\rm \Delta(BP-RP)}$ $\approx$  0.018 (before and after applying the T$_{\rm eff}$-offset, respectively). We therefore used the shifted stellar temperatures (points from the right-hand panel) in the projection on the PARSEC isochrones to estimate the final stellar ages. \par

Moreover, the estimated distance from \emph{Gaia} DR2 parallaxes was taken into account to compute the distance modulus
\ of each star in order to calculate its absolute magnitude. We firstly estimated the Galactic extinction for our sample by implementing the 3D dust maps from \citet{green2018}, and also \citet{schlegel1998} maps with the proposed correction by \citet[][see Equation 24]{sharma2014} in order to not overestimate the reddening. We calculated a negligible extinction (E(B-V) $\lesssim$ 0.025~mag) for most of the observed stars, while some cases showed significant values but with uncertainties as high as the estimated correction. The very low derived extinctions are consistent with the fact that our sample is in the solar vicinity (d < 300 pc). As a consequence, we decided not to apply any extinction correction, as they have a negligible impact on our results. Furthermore, we did not adopt any prior either for the Galaxy model or for the age as a function of other stellar parameters such as [M/H] or [Mg/Fe]. We assumed a uniform star formation history to avoid prioritising a particular formation epoch.  \par

Finally, we excluded the stars whose mean, median, and mode age PDF values differed  from one another by more than 2~Gyr, leading to a total of 366 stars with a very reliable age estimate. The adopted age for each star was the one derived by the mean of the PDF. The age distribution for our data sample ranges from 2.5 to 13.5~Gyr, with an average relative standard deviation of $\sigma~\sim$~20~\% around the mean PDF value. The absence of stars younger than 2.5~Gyr may be due to a possible bias inherent to the sample.

\subsection{Kinematic and orbital properties} \label{orbits}

We estimated the orbital parameters with the \emph{Gaia} DR2 astrometric positions and proper motions, the calculated distances by \citet{bailer-jones2018}, and the spectroscopic radial velocities determined  by the AMBRE analysis procedure \citep[see][]{worley2012, DePascale2014}. For this purpose, we used the Python code \texttt{galpy} \citep{bovy2015}, together with the \texttt{MWpotential2014}: a Milky-Way-like gravitational potential that is the sum of a power-law density profile for the bulge (power-law  exponent of -1.8 and a cut-off radius of 1.9 kpc), a Miyamoto-Nagai potential for the disc \citep{miyamoto1975}, and a Navarro-Frenk-White potential to model the halo \citep{NFW1997}. The orbits were integrated over 10 Gyr in order to evaluate the pericentres (R$_{\rm per}$) and the apocentres (R$_{\rm apo}$), as well as the maximum heights above the Galactic plane (|z$_{\rm max}$|). As we analysed a local stellar sample in the solar neighbourhood (large parallaxes: 3~<~$\varpi$~<~50~mas), the parallax bias found in the \emph{Gaia} DR2 data, for example the global shift of $\delta_\varpi$~= -0.029 mas \citep{lindegren2018} or $\delta_\varpi$~=~-0.054~mas by later studies \citep{schonrich2019, graczyk2019}, has a negligible effect on the estimated kinematic and orbital quantities. \par

Figure \ref{Fig:hist_ALL} shows the distribution of the main stellar properties of our final sample, consisting of 366 MSTO stars. These are Galactic disc stars, describing relatively circular prograde orbits (0 < e $\lesssim$ 0.4) close to the Galactic plane (|z$_{\rm max}$| $\lesssim$ 1~kpc), and are evenly distributed in age from $\sim$~2.5 to 13.5~Gyr. We find a significant fraction of stars that seem to be passing through our local limited sample on the present day, with R$_{\rm per}$ values lower than 6 kpc, and also some of them reaching  R$_{\rm apo}$ values of greater than 10 kpc from the Galactic centre. We used the guiding centre radius (R$\rm _g$) as an estimate of the current value of the Galactocentric radius (R$_{\rm GC}$), because it is a useful tracer of the radial migration of the different stellar populations in the Galactic disc \citep[e.g.][]{binney2007,georges2017}. For this work, R$\rm _g$ was calculated as the average between the pericentre and the apocentre of the orbit of the star: \(\rm R_g = (R_{per} + R_{apo}) / 2 \). The sample is well distributed in Galactocentric R$_{\rm g}$ from 4 to 11~kpc.
\par

In order to estimate the error in R$\rm _g$, we re-calculated the orbits using the upper and lower distance limit derived by \citet{bailer-jones2018}. The comparison shows a negligible uncertainty of less than 1 parsec. We also calculated the orbits using other potentials: the \texttt{MWpotential2014} adding a bar (\texttt{DehnenBarPotential}) on the one hand, and the potential presented by \citet{McMillan2017} on the other. For both cases, the differences in the R$_{\rm g}$ value were around 0.2~kpc with respect to the initial calculated ones. Our results are robust against the choice of the Galactic potential model and the uncertainty in the parameters. 

\section{Radial chemical trends and stellar migration} \label{results_1}

In this section, we first present an exploration of the present-day distribution of [Mg/Fe] and [M/H] in the thin disc as a function of Galactocentric position and age. We then present our analysis of the impact of radial migration in the solar neighbourhood.

\subsection{Definition of the thin disc} \label{disc_distinction}

Here we adopt a chemical definition of the thin disc, based on the [Mg/Fe] content in a given [M/H] bin.

Figure \ref{Fig:THICKvsTHIN} illustrates the chemical separation in the [Mg/Fe]--[M/H] plane for our working sample, classifying the stars into high- and low-[Mg/Fe] sequences. First of all, we selected stars older than 12~Gyr. According to previous works in the literature \citep[c.f.][]{fuhrmann2011,haywood2013,hayden2017}, this would trace the old (thick) disc population. These stars, highlighted by orange squares in Fig. \ref{Fig:THICKvsTHIN}, are mostly [Mg/Fe]-enhanced ([Mg/Fe] $\gtrsim$ + 0.2 dex), but span a wide range in [M/H], reaching the solar values. We used the lower [Mg/Fe]-bound~fit\footnote{[Mg/Fe] = -0.3 · [M/H] + 0.045 (dex)} to these stars to chemically define the thick (blue stars in Fig.~\ref{Fig:THICKvsTHIN}) and the thin disc (red points in Fig.~\ref{Fig:THICKvsTHIN}). This line clearly separates the two chemical sequences at low metallicities, yet the extrapolation to the metal-rich tail ([M/H]~$\textgreater$~-0.3~dex) is rather arbitrary and the classified thick disc, also called the $\alpha$-rich metal-rich population in previous works \citep[e.g.][]{vardan2012}, has an \emph{ad~hoc} assignment that needs to be analysed in detail. \par


\subsubsection*{Distinct metal-rich populations?} \label{metalrich_distinction}

The difficulty in   separating the thin and thick disc stars in terms of chemical abundances at high metallicities is a matter of debate in the literature. On the one hand, \citet{vardan2012} and \citet{sarunas2017} found a gap in metallicity ([Fe/H]~$\approx$ -0.2~dex) for the thick disc population in a sample of dwarf stars  from HARPS data. These authors used this gap to chemically define the thick disc and the $\alpha$-rich metal-rich sequences separately. \citet{vardan2012} showed that high-$\alpha$ metal-rich stars were on average older than thin disc stars, but with similar kinematics and orbits to the thin disc population. On the other hand, \citet{hayden2015} and \citet{buder2019} (for two independent samples of giants from APOGEE and dwarfs from GALAH DR2 data, respectively) found a continuous evolution for the thick disc sequence, describing an independent track from the thin disc up to supersolar metallicities. \par

\begin{figure}
\centering
\includegraphics[height=65mm, width=0.48\textwidth]{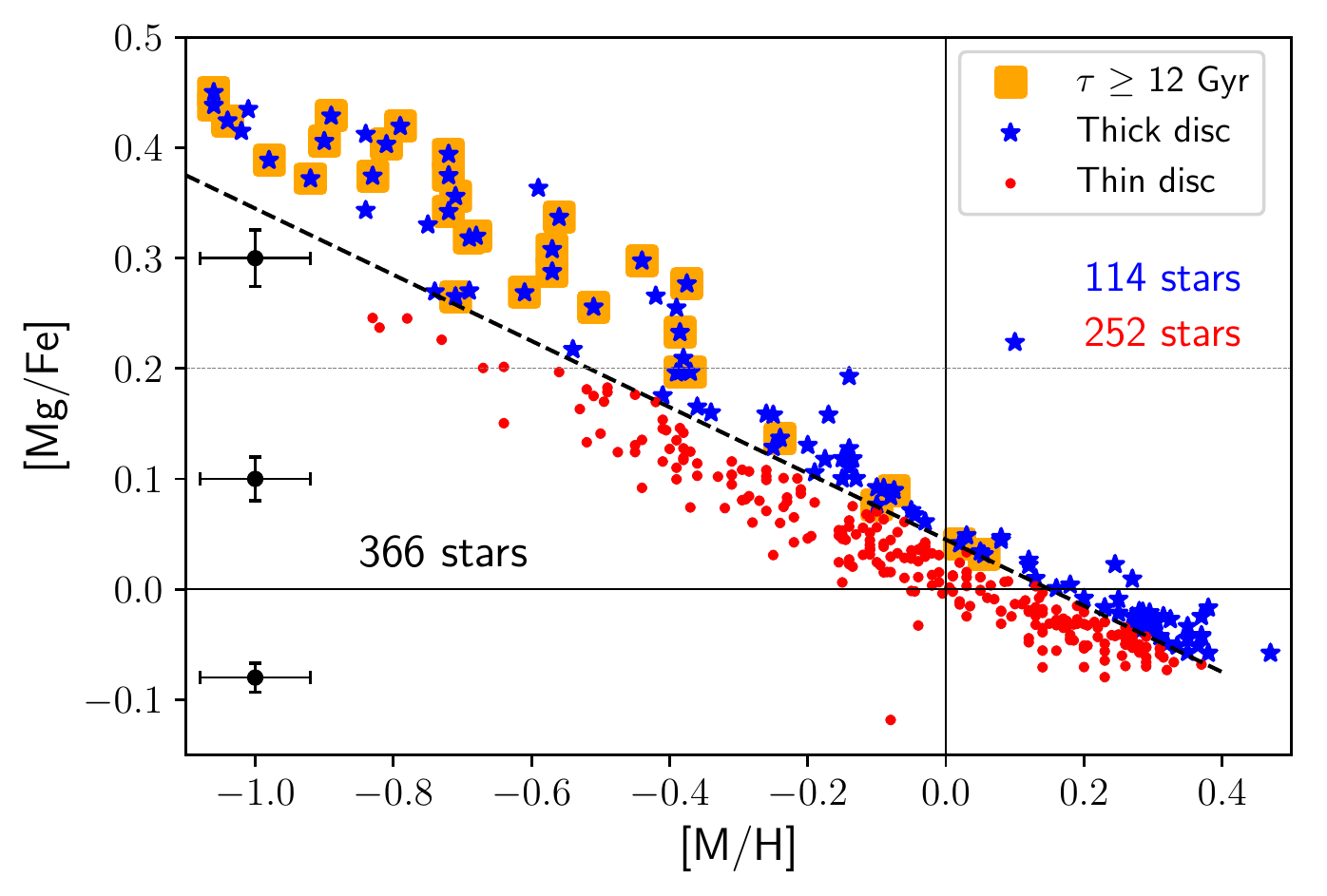}
\caption{[Mg/Fe] vs. [M/H] for our working sample. The stars with ages older than 12 Gyr are highlighted with orange squares. The black dashed line defines the thin(red circles)--thick(blue stars) disc chemical separation. The mean estimated errors are represented on the left-hand side for three different intervals in [Mg/Fe].}
\label{Fig:THICKvsTHIN}
\end{figure}


\begin{figure*}
\centering
\includegraphics[height=90mm, width=\textwidth]{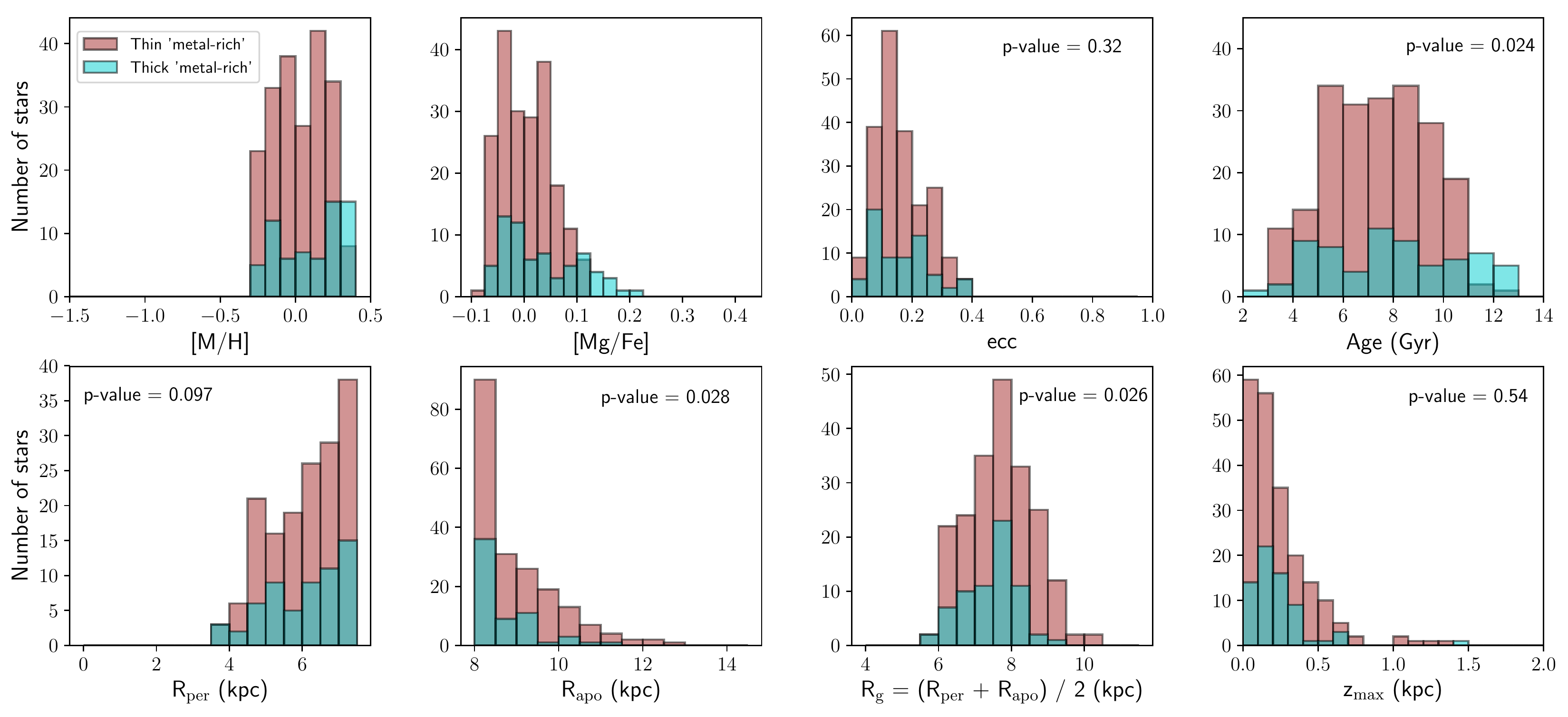}
\caption{Same stellar properties as Fig. \ref{Fig:hist_ALL}, separated into the chemically defined thin (brown) and thick (blue) disc stellar populations for the metal-rich subsample ([M/H] > -0.3~dex). The p-values of the two-sample Kolmogorov-Smirnov tests are reported for each parameter distribution.}
\label{Fig:hist_DIST_rich}
\end{figure*}

\begin{figure*}
\centering
\includegraphics[height=90mm, width=\textwidth]{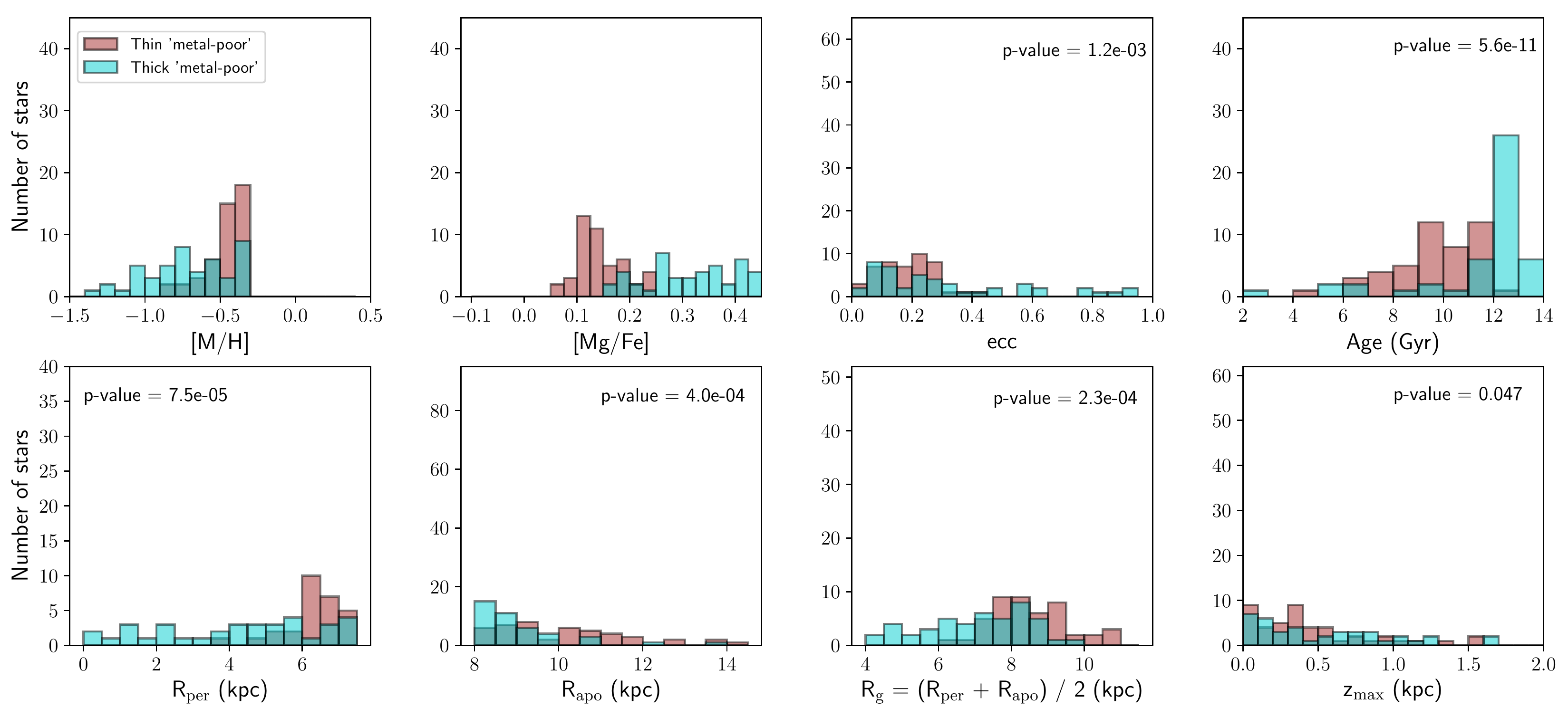}
\caption{Same as Fig. \ref{Fig:hist_DIST_rich} but for the metal-poor subsample ([M/H] $\leq$ -0.3~dex), with the respective p-values of the two-sample Kolmogorov-Smirnov tests.}
\label{Fig:hist_DIST_poor}
\end{figure*}

Moreover, due to possible ignored uncertainties in our abundance estimates \citep{SantosPeral2020}, each sample is expected to contain a fraction of contamination from the other sample in the metal-rich regime ([M/H] > -0.3~dex). To evaluate the advantages of a separate treatment of the thick disc metal-rich population with respect to the thin disc one, we assessed the possible differences between the classified high- and low-[Mg/Fe] disc sequence populations in the metal-rich regime. 
Figure \ref{Fig:hist_DIST_rich} shows the chemodynamical properties of the high- (blue) and low-[Mg/Fe] (brown) disc metal-rich stars ([M/H]~> -0.3~dex). Some of the high-[Mg/Fe] disc metal-rich stars show low eccentric orbits, young ages, and are close to the plane (low-z$_{\rm max}$), similarly to the metal-rich low-[Mg/Fe] population. The Kolmogorov-Smirnov test between the two samples does not allow us to reject that these high metallicity stars are truly different stellar populations. This is particularly true for the z$_{\rm max}$ and the eccentricity distributions, while the p-value is smaller for the radial and age distribution. 
\par 

For comparison, Fig.~\ref{Fig:hist_DIST_poor} illustrates the same analysis for the metal-poor stellar subsample ([M/H]~$\leq$~-0.3~dex). Besides the clear [Mg/Fe] distinction observed in Fig.~\ref{Fig:THICKvsTHIN} at this metallicity regime, it seems reasonable to interpret the reported p-values as evidence in favour of two different stellar populations (p-values~<~10$^{-3}$ for the orbital parameter distributions: eccentricity, R$_{\rm per}$, R$_{\rm apo}$,  and R$_{\rm g}$). The high-[Mg/Fe] metal-poor population clearly shows a more centrally concentrated distribution (reaching the innermost regions: R$_{\rm per}$~$\sim$~0-7~kpc and R$_{\rm g}$ down to 4~kpc from the Galactic centre) and presents a higher eccentricity tail. On the contrary, the low-[Mg/Fe] metal-poor population reaches the outer parts of the Galactic disc (up to R$_{\rm apo}$~$\sim$~14~kpc and R$_{\rm g}$~$\sim$~10~kpc) and only describes circular orbits (e $\lesssim$ 0.3). \par

In conclusion, we only decided to apply the chemical criterion described in Fig. \ref{Fig:THICKvsTHIN} in the metal-poor regime ([M/H] $\leq$ -0.3~dex) in order to minimise the thick disc contamination in the study of the thin disc properties. Nevertheless, the similarities observed in the metal-rich regime ([M/H] > -0.3~dex) do not support a thin--thick disc separation at high metallicities. As a consequence, we consider all the metal-rich stars as part of the thin disc population in the following gradient and radial migration study, and a global consideration of the entire disc population is adopted for the disc evolution analysis in Section~\ref{results_2}. 


\subsection{Present-day chemical abundance gradients} \label{gradients}

Table \ref{table:gradient} shows the radial gradients of [M/H] and [Mg/Fe] for the thin disc stars, assuming R$\rm _g$ as their true position in the Galaxy. The values of the slope and the uncertainties are reported in Table \ref{table:gradient}, and come from a Theil-Sen regression model. \par


We find a negative gradient of -0.099 $\pm$ 0.031 dex~kpc$^{-1}$ for [M/H] and a positive gradient of +0.023 $\pm$ 0.009 dex~kpc$^{-1}$ for the [Mg/Fe] abundance. Both chemical gradients are flatter for young stars ($\leq$~6~Gyr), although the differences are within the slope uncertainties. 
In contrast, for the same stellar subsample, the \citet{sarunas2017} [Mg/Fe] abundances lead to a shallower gradient: +0.004 $\pm$ 0.007 dex~kpc$^{-1}$. \par

The [Mg/Fe] abundances from \citet{SantosPeral2020}, obtained with a careful treatment of the spectral continuum placement, show a decreasing trend in the [Mg/Fe] abundance even at supersolar metallicites. In contrast,  previous observational studies of the solar neighbourhood observed a flattened trend \citep[e.g.][]{vardan2012, sarunas2017}. As a consequence, it seems that the reported improvement in the [Mg/Fe] abundance precision for the metal-rich disc implies a significant change in the radial gradient measurement, showing steeper slopes. As an additional test, we also explored the possible influence of a thin--thick disc misclassification (see Sect.~\ref{metalrich_distinction}) by not including in our thin disc radial gradient the measurements of the high-[Mg/Fe] metal-rich stars, finding similar results: d[Mg/Fe]/dR$_{\rm g}$~=~+0.025 $\pm$ 0.009 dex~kpc$^{-1}$ (whole sample), +0.020 $\pm$ 0.015~dex~kpc$^{-1}$ (young stars), and +0.028~$\pm$~0.010~dex~kpc$^{-1}$ (old stars). Therefore, the observed steeper [Mg/Fe] gradient is not affected by the possible thin--thick disc contamination, and seems to be a direct result of the newly derived [Mg/Fe] abundances.

\begin{table*}
\centering
\caption{Radial abundance gradients (dex~kpc$^{-1}$) in the range 6 $\leq$ R$\rm _g$ $\leq$ 11 kpc for Galactic thin disc stars.} 
\begin{tabular}{cccccccc}
\hline
\hline
\\[\dimexpr-\normalbaselineskip+2pt]
& \( \rm \frac{d[M/H]}{dRg} \) & \(\rm  \frac{d[Mg/Fe]}{dRg} \)  & \multicolumn{1}{c}{\(\rm  \frac{d[Mg/Fe] ^\dagger}{dRg} \)} 
\\
\\[\dimexpr-\normalbaselineskip+2pt]
\hline
\\[\dimexpr-\normalbaselineskip+2pt]
Whole sample & -0.099 $\pm$ 0.031 & +0.023 $\pm$ 0.009 & \multicolumn{1}{c}{+0.004 $\pm$ 0.007} 
\\
\\[\dimexpr-\normalbaselineskip+2pt]
Young ($\leq$ 6 Gyr) & -0.088 $\pm$ 0.063 & +0.021 $\pm$ 0.015 & \multicolumn{1}{c}{-0.007 $\pm$ 0.011} 
\\
\\[\dimexpr-\normalbaselineskip+2pt]
Old (> 6 Gyr) & -0.106 $\pm$ 0.035 & +0.025 $\pm$ 0.010 & \multicolumn{1}{c}{+0.010 $\pm$ 0.007} 
\\
\hline
\\[\dimexpr-\normalbaselineskip+2pt]
\hline
\\[\dimexpr-\normalbaselineskip+2pt]
\multicolumn{4}{l}{\scriptsize $\dagger$ Abundances from \citet{sarunas2017}}
\end{tabular}
\label{table:gradient}
\end{table*}

\begin{table*}
\centering
\caption{Radial abundance gradients (dex~kpc$^{-1}$) in the Galactic thin disc from the literature.}
\begin{tabular}{ccccp{3.8cm}c}
\hline
\hline
\\[\dimexpr-\normalbaselineskip+2pt]
Abundance & Value & Work & Tracer & \hspace{0.5cm}Thin disc definition & Database \\
\\[\dimexpr-\normalbaselineskip+2pt]
\hline
\hline
\\[\dimexpr-\normalbaselineskip+2pt]
\( \rm \frac{d[Fe/H]}{dR_{\rm GC}} \) &  -0.066 $\pm$ 0.037 & \citet{cheng2012} & Field dwarf stars & \hspace{0.3cm} 0.15< |z| < 0.25 kpc\par \hspace{0.4cm} (6 < R$_{\rm GC}$ < 16 kpc) & SEGUE \\
\\[\dimexpr-\normalbaselineskip+2pt]
\( \rm \frac{d[Fe/H]}{dR_{\rm g}} \) &  -0.065 $\pm$ 0.003 & \citet{boeche2013} & Field dwarf stars & \hspace{0.7cm} |z$_{\rm max}$| < 0.4 kpc \par \hspace{0.4cm} (4.5 < R$_{\rm g}$ < 9.5 kpc) & RAVE \\
\\[\dimexpr-\normalbaselineskip+2pt]
\( \rm \frac{d[M/H]}{dR_{\rm GC}} \) &  -0.090 $\pm$ 0.002 & \citet{hayden2014} & Field giant stars &  Chemical + |z| < 0.25 kpc \par \hspace{0.4cm} (5 < R$_{\rm GC}$ < 15 kpc) & APOGEE \\
\\[\dimexpr-\normalbaselineskip+2pt]
\( \rm \frac{d[Fe/H]}{dR_{\rm g}} \) &  -0.074 $\pm$ 0.010 & \citet{anders2014} & Field giant stars & \hspace{0.7cm} |z$_{\rm max}$| < 0.4 kpc \par \hspace{0.5cm} (6 < R$_{\rm g}$ < 11 kpc) & APOGEE Gold \\
\\[\dimexpr-\normalbaselineskip+2pt]
\( \rm \frac{d[Fe/H]}{dR_{\rm GC}} \) &  -0.068 $\pm$ 0.016 & \citet{bergemann2014} & Field stars & \hspace{0.8cm} |z| < 0.3 kpc \par \hspace{0.4cm} (6 < R$_{\rm GC}$ < 9.5 kpc) & GES UVES \\
\\[\dimexpr-\normalbaselineskip+2pt]
\( \rm \frac{d[Fe/H]}{dR_{\rm GC}} \) &  -0.045 $\pm$ 0.012 & \citet{sarunas2014} & Field stars & Chemical + |z| < 0.608 kpc \par \hspace{0.4cm} (4 < R$_{\rm GC}$ < 12 kpc) & GES GIRAFFE \\
\\[\dimexpr-\normalbaselineskip+2pt]
\( \rm \frac{d[Fe/H]}{dR_{\rm GC}} \) &  -0.058 $\pm$ 0.008 & \citet{alejandra2014} & Field stars & Chemical + |z| < 0.7 kpc \par \hspace{0.4cm} (5 < R$_{\rm GC}$ < 11 kpc) & GES GIRAFFE \\
\\[\dimexpr-\normalbaselineskip+2pt]
 &  -0.053 $\pm$ 0.029 &  & \hspace{0.95cm} $\leq$ 0.8 Gyr & & \\
\( \rm \frac{d[Fe/H]}{dR_{\rm GC}} \) & -0.094 $\pm$ 0.008 & \citet{magrini2009} & OCs: 0.8 - 4 Gyr & \hspace{0.5cm} (7 < R$_{\rm GC}$ < 12 kpc) & \\
&  -0.091 $\pm$ 0.060 &  &  \hspace{0.95cm} 4 - 11 Gyr & &  \\
\\[\dimexpr-\normalbaselineskip+2pt]
\( \rm \frac{d[Fe/H]}{dR_{\rm GC}} \) & -0.090 $\pm$ 0.030 & \citet{frinchaboy2013} & OCs & \hspace{0.4cm} (7.9 < R$_{\rm GC}$ < 14 kpc) & OCCAM-APOGEE \\
\\[\dimexpr-\normalbaselineskip+2pt]
\( \rm \frac{d[Fe/H]}{dR_{\rm GC}} \) & -0.068 $\pm$ 0.004 & \citet{donor2020} & OCs & \hspace{0.4cm} (6 < R$_{\rm GC}$ < 13.9 kpc) & OCCAM-APOGEE \\
\\[\dimexpr-\normalbaselineskip+2pt]
\( \rm \frac{d[Fe/H]}{dR_{\rm GC}} \) & -0.060 $\pm$ 0.002 & \citet{genovali2014} & Cepheids & \hspace{0.5cm} (5 < R$_{\rm GC}$ < 19 kpc) & \\
\\[\dimexpr-\normalbaselineskip+2pt]
\hline
\\[\dimexpr-\normalbaselineskip+4pt]
\( \rm \frac{d[Mg/Fe]}{dR_{\rm g}} \) &  -0.009 $\pm$ 0.002 & \citet{boeche2013} & Field dwarf stars & \hspace{0.7cm} |z$_{\rm max}$| < 0.4 kpc \par \hspace{0.4cm} (4.5 < R$_{\rm g}$ < 9.5 kpc) & RAVE \\
\\[\dimexpr-\normalbaselineskip+2pt]
\( \rm \frac{d[Mg/Fe]}{dR_{\rm GC}} \) &  +0.021 $\pm$ 0.016 & \citet{bergemann2014} & Field stars & \hspace{0.8cm} |z| < 0.3 kpc \par \hspace{0.4cm} (6 < R$_{\rm GC}$ < 9.5 kpc) & GES UVES \\
\\[\dimexpr-\normalbaselineskip+2pt]
\( \rm \frac{d[Mg/M]}{dR_{\rm GC}} \) &  +0.009 $\pm$ 0.003 & \citet{sarunas2014} & Field stars & Chemical + |z| < 0.608 kpc \par \hspace{0.4cm} (4 < R$_{\rm GC}$ < 12 kpc) & GES GIRAFFE \\
\\[\dimexpr-\normalbaselineskip+2pt]
\( \rm \frac{d[Mg/Fe]}{dR_{\rm GC}} \) &  +0.012 $\pm$ 0.002 & \citet{alejandra2014} & Field stars & Chemical + |z| < 0.7 kpc \par \hspace{0.4cm} (5 < R$_{\rm GC}$ < 11 kpc) & GES GIRAFFE \\
\\[\dimexpr-\normalbaselineskip+2pt]
\( \rm \frac{d[Mg/Fe]}{dR_{\rm GC}} \) &  +0.013 $\pm$ 0.006 & \citet{sarunas2019} & Field dwarf stars & \hspace{0.9cm} Kinematical \par \hspace{0.4cm} (6 < R$_{\rm GC}$ < 10 kpc) & VUES \\
\\[\dimexpr-\normalbaselineskip+2pt]
\( \rm \frac{d[Mg/Fe]}{dR_{\rm GC}} \) & +0.013 $\pm$ 0.003 & \citet{genovali2015} & Cepheids & \hspace{0.2cm} (4.1 < R$_{\rm GC}$ < 18.4 kpc) & UVES \\
\\[\dimexpr-\normalbaselineskip+2pt]
\( \rm \frac{d[Mg/Fe]}{dR_{\rm GC}} \) & +0.009 $\pm$ 0.001 & \citet{donor2020} & OCs & \hspace{0.3cm} (6 < R$_{\rm GC}$ < 13.9 kpc) & OCCAM-APOGEE \\
\\[\dimexpr-\normalbaselineskip+2pt]
\hline
\\[\dimexpr-\normalbaselineskip+2pt]
\hline
\end{tabular}
\label{table:gradient3}
\end{table*}

\subsubsection*{Comparison with literature studies}

The radial metallicity gradient value in the Galactic disc, along with its evolution with time, are still debated in the literature, with negative slopes ranging from -0.04 to~-0.1 dex~kpc$^{-1}$ for the analysis of different Galactic populations: planetary nebulae (PNe), HII regions, open clusters, variable (Cepheids) and field stars. Our measured value is therefore in agreement within the literature range (see Table \ref{table:gradient3}). It is worth noting that the gradients shown here correspond to works with different target selections and implemented methodologies to estimate abundances and distances. Depending on the work, the metallicity value was measured globally ([M/H]) or considering only iron lines ([Fe/H]), and the radial distance estimate comes from either the guiding centre radius of the stellar orbit (R$_{\rm g}$) or the Galactocentric radius of the star (R$_{\rm GC}$). \par

First, Table~\ref{table:gradient3} shows how our metallicity gradient estimation is qualitatively in good agreement with previous analyses of field stars from different Galactic surveys. Compatible results were obtained using data from SEGUE  \citep[R $\sim$ 2000,][]{cheng2012}, RAVE \citep[R $\sim$ 7500,][]{boeche2013}, APOGEE \citep[R $\sim$ 22500,][]{hayden2014, anders2014}, and the Gaia-ESO Survey (GES) GIRAFFE \citep[R $\sim$ 20000,][]{sarunas2014, alejandra2014} and UVES data \citep[R $\sim$ 47000,][]{bergemann2014}.

Additionally, stellar open clusters (OCs) are considered a unique tool with which to study the time evolution of the radial metallicity gradients, because of their accurate derived ages and Galactocentric distances, covering wide ranges along the Galactic disc \citep[c.f.][OCs ages from $\sim$30 Myr to 11 Gyr]{magrini2009}. Our measurement shows good agreement with the one presented by \citet{magrini2009} for OCs older than 0.8 Gyr, and also with the overall gradient found by \citet{frinchaboy2013} from the OCCAM survey using APOGEE data, which was recently updated by \citet{donor2020}. \par

Furthermore, Cepheid variable stars provide the present-day abundance gradients of the Galactic disc because they are massive stars younger than 300 Myr, and are commonly used as good distance indicators and as chemical tracers of the interstellar medium (ISM) abundance \citep[c.f.][and references therein]{genovali2014}. For a homogeneous sample of Galactic Cepheids observed at high spectral resolution (R $\sim$ 38000), \citet{genovali2014} derived a consistent estimate for the present-day ISM abundance gradient.  \par

On the other hand, the observed positive gradient in the [Mg/Fe] abundance with radius is in close agreement with the results from \citet[][]{bergemann2014}. However, shallower positive gradients in [Mg/Fe] are commonly reported within the literature \citep[e.g.][a recent analysis using the high-resolution VUES spectrograph, R $\sim$ 60000]{sarunas2014, alejandra2014, sarunas2019}, even finding negative values \citep[][]{boeche2013}. Additionally, \cite{genovali2015} and \citet{donor2020} reported a similar slope resulting from an analysis of Galactic Cepheids and open clusters, respectively, over the Galactic disc. \par

As a consequence, our radial [Mg/Fe] gradient is not only steeper than the one estimated from \citet{sarunas2017} abundances (see Table \ref{table:gradient}), but is also steeper than other previous measurements in the literature. Once again, this is a major consequence of the \citet{SantosPeral2020} improvement of the  data for metal-rich stars. However, we point out that \citet[][]{perdigon2020} report a flatter radial gradient for sulphur (another $\alpha$-element) equal to +0.012 dex~kpc$^{-1}$ for AMBRE stars and adopting a similar methodology to ours.

\subsection{Radial migration} \label{migration}

The stars in the Galactic disc are very likely to be scattered by resonances with the spiral arms or by giant molecular clouds that could increase the radial oscillation amplitudes around their R$\rm _g$ (blurring), or change the angular momentum of the orbit (churning) \citep{kalnajs1972, lacey1984, grenon1989, grenon1999, sellwood2002, schonrich2009, minchev2014b,georges2015a}. On the one hand, stars with more eccentric and inclined orbits can belong to Galactic regions far from the Sun but reach the solar neighbourhood via blurring. However, as shown in Fig. \ref{Fig:hist_ALL}, 50 \% of the stars of our selected sample are on very circular orbits with e~<~0.15. Therefore, the observed stars are either expected to have been born locally or to have increased (decreased) their angular momentum via churning, increasing (decreasing) their guiding centre radius~(R$\rm _g$) without changing their orbit eccentricity. \par

The impact of radial migration over a whole observed stellar sample is a complex, unsolved problem. In particular, its final effect on the chemical evolution and the metallicity distribution function in the solar vicinity is very uncertain, and has been suggested to be negligible \citep{spitoni2015, halle2018, vincenzo2020, khoperskov2020}. The unique observational constraint consists in the presence of too metal-rich stars in the solar annulus that cannot be justified by chemical evolution models without an exchange of matter between different radial annuli in the Galactic disc \citep{grisoni2017}. Under the assumption that a negative radial metallicity gradient has been present in the Galaxy since the thin disc formation epoch \citep{roskar2008, magrini2009, schonrich2017, minchev2018}, and given a present-day ISM abundance in the solar vicinity of around [M/H] $\sim$ 0.0 dex, the observed super-metal-rich stars (SMR; [M/H] $\gtrsim$ +0.1 dex) should have been formed in the inner disc regions \citep{minchev2013, georges2015a, hayden2020}. \par 

These assumptions have been taken into account in the following analysis, although some uncertainties remain. For instance, the metallicity of forming stars in the solar vicinity is still a matter of debate, with some recent studies questioning its solar nature \citep[e.g. open cluster analysis by][]{baratella2020, spina2021}, and also \citet{delgadomena2019} found [Fe/H]~$\textgreater$~0 (with a mean near 0.1 dex) for field stars younger than 1~Gyr. In addition, based on the observed cosmic dispersion in metallicity in the local Universe \citep[$\sim$~0.05~dex, e.g.][]{mannucci2010} and the observational error of our measurements (0.04~dex; see Section~\ref{harps}), the presence of stars with metallicity in the range 0.1-0.2 could also be compatible with a chemical evolution with no radial migration in the solar annulus. \par 


Following the procedure described in \citet{hayden2020}, we estimated the minimum required eccentricity of a star at a given metallicity to reach the solar neighbourhood exclusively due to blurring. For this purpose, we assumed the apocentre of the  orbit to be the measured present-day position. Additionally, assuming a given radial ISM metallicity gradient and fixing the local ISM abundance to (R$_\odot$, [M/H]) = (8.0 kpc, 0.0 dex), it is possible to estimate the birth radius of the star from its observed present-day [M/H]. In this framework, we selected the churned candidates through the following required minimum eccentricity relation:

\begin{equation} \label{eq:ecc}
\centering
\rm ecc^{\star}([M/H]) \geq \frac{R}{R_{birth}([M/H])} - 1
.\end{equation}

In this relation, R is the present-day  star position, which is assumed to lie between 7.7 and 8.3 kpc because our sample is located within 300 pc of the Sun, and R$_{\rm birth}$ is the estimated birth radius of the star given its metallicity [M/H], but also assuming an ISM gradient. \par 

The evolution with time of the ISM radial metallicity gradient is also debated in the literature for Galactic chemical evolution models and cosmological simulations: some models predict a time invariant gradient \citep[e.g.][]{magrini2009, gibson2013}, while others predict a steepening of the ISM metallicity gradient with time \citep[e.g.][]{chiappini2001, schonrich2017}, or a flattening with time \citep[e.g.][]{prantzos1999, hou2000,roskar2008, pilkington2012, minchev2018}. For that reason, we decided to consider different ISM gradient values in the R$_{\rm birth}$ estimate for the SMR stars, and study the influence of these assumptions on the conclusions. 
\par

\begin{figure}
\centering
\includegraphics[height=70mm, width=0.48\textwidth]{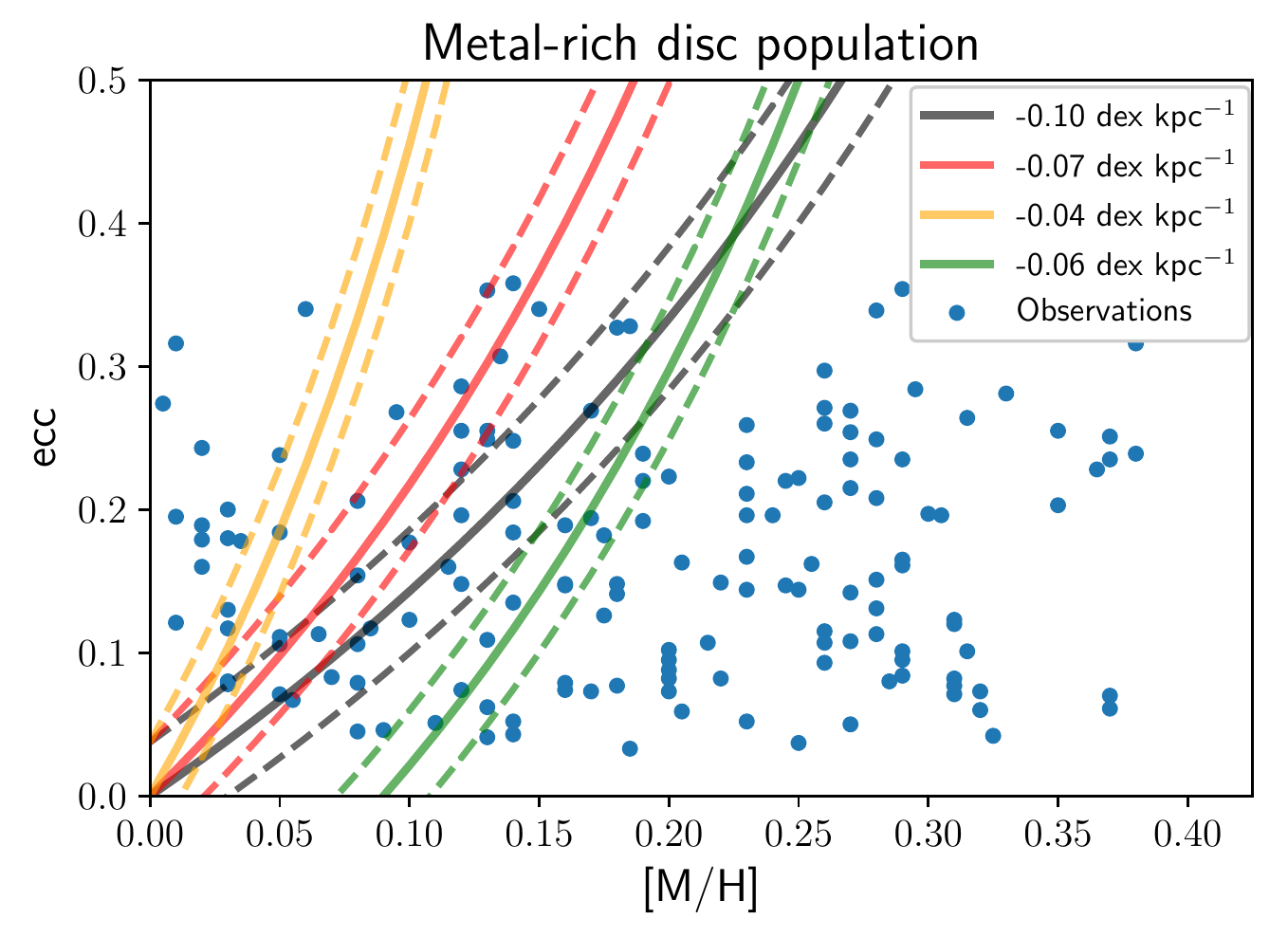}
\caption{Distribution of the orbital eccentricities as a function of [M/H] for the thin disc metal-rich sample. The solid (dashed) lines indicate the required eccentricity to reach R~=~8~kpc ($\pm$ 300 pc) without the need for churning (see~Eq.~\ref{eq:ecc}). For a fixed zero point, ISM$_{\rm [M/H]}$(R$_\odot$)~=~0.0, we studied three different ISM metallicity gradients (black, red, and orange). The green curve corresponds to the gradient from \citet{genovali2014}. The stars to the right require churning to reach the solar neighbourhood.}
\label{Fig:MIGRATED_candidates}
\end{figure}

\begin{figure*}
\centering
\includegraphics[height=50mm, width=\textwidth]{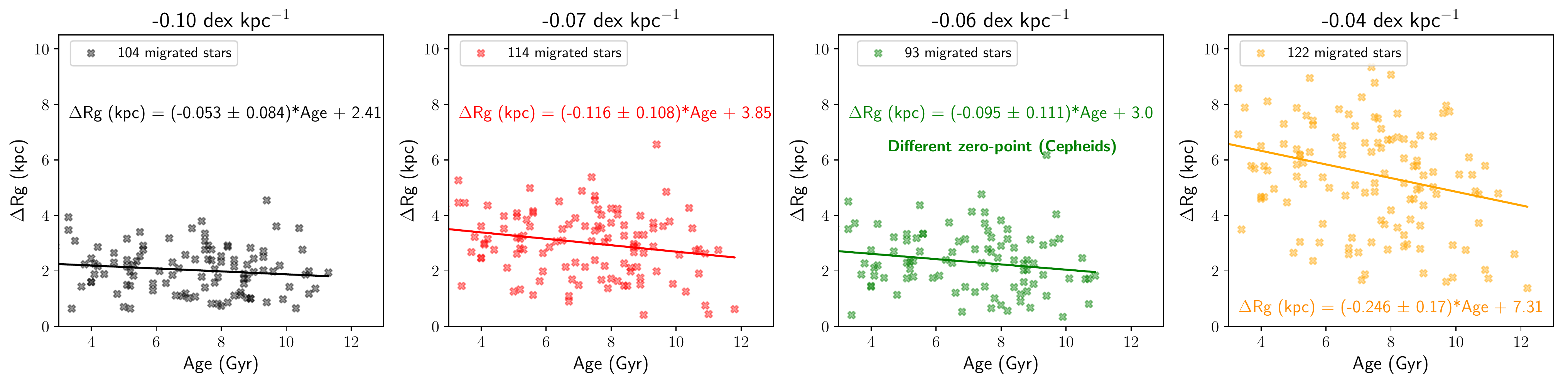}
\caption{Estimated covered distance $\Delta$R$\rm _g$ (R$\rm _g$ - R$_{\rm birth}$) vs. stellar age for the selected churned SMR stars ([M/H]~$\textgreater$~0.1) for different ISM metallicity gradients (see Fig.~\ref{Fig:MIGRATED_candidates} and titles of the plots).} 
\label{Fig:migrated_distance}
\end{figure*}

\begin{figure*}
\centering
\includegraphics[height=50mm, width=\textwidth]{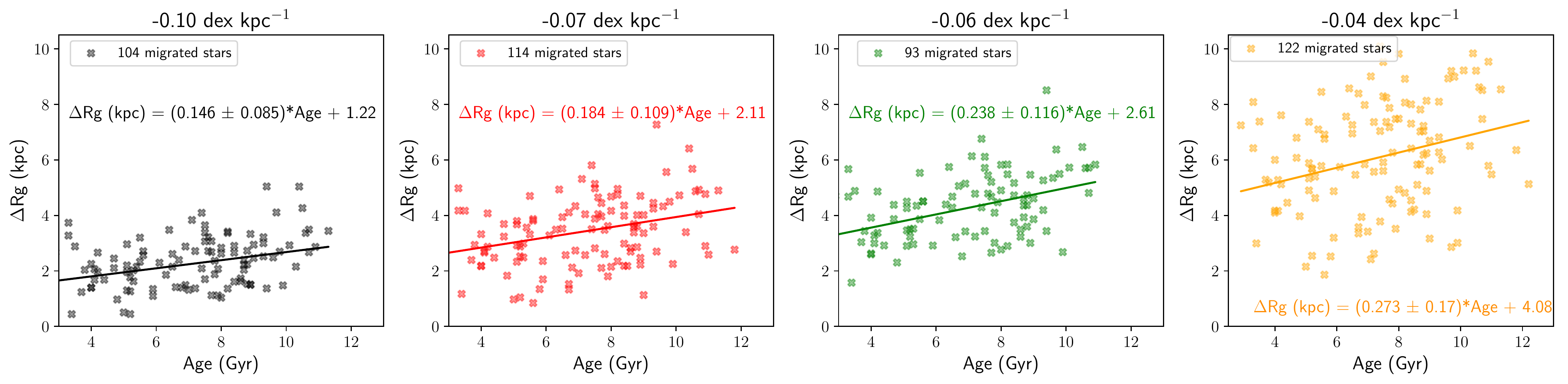}
\caption{Same as Fig. \ref{Fig:migrated_distance}, but applying a different zero point as a function of the stellar age (ISM$_{\rm [M/H]}$(R$_\odot$,$\tau$)) in the R$_{\rm birth}$ estimate (see Sect~\ref{migration}).}
\label{Fig:migrated_distance_zeropoint}
\end{figure*}

\begin{figure}
\centering
\includegraphics[height=65mm, width=0.4\textwidth]{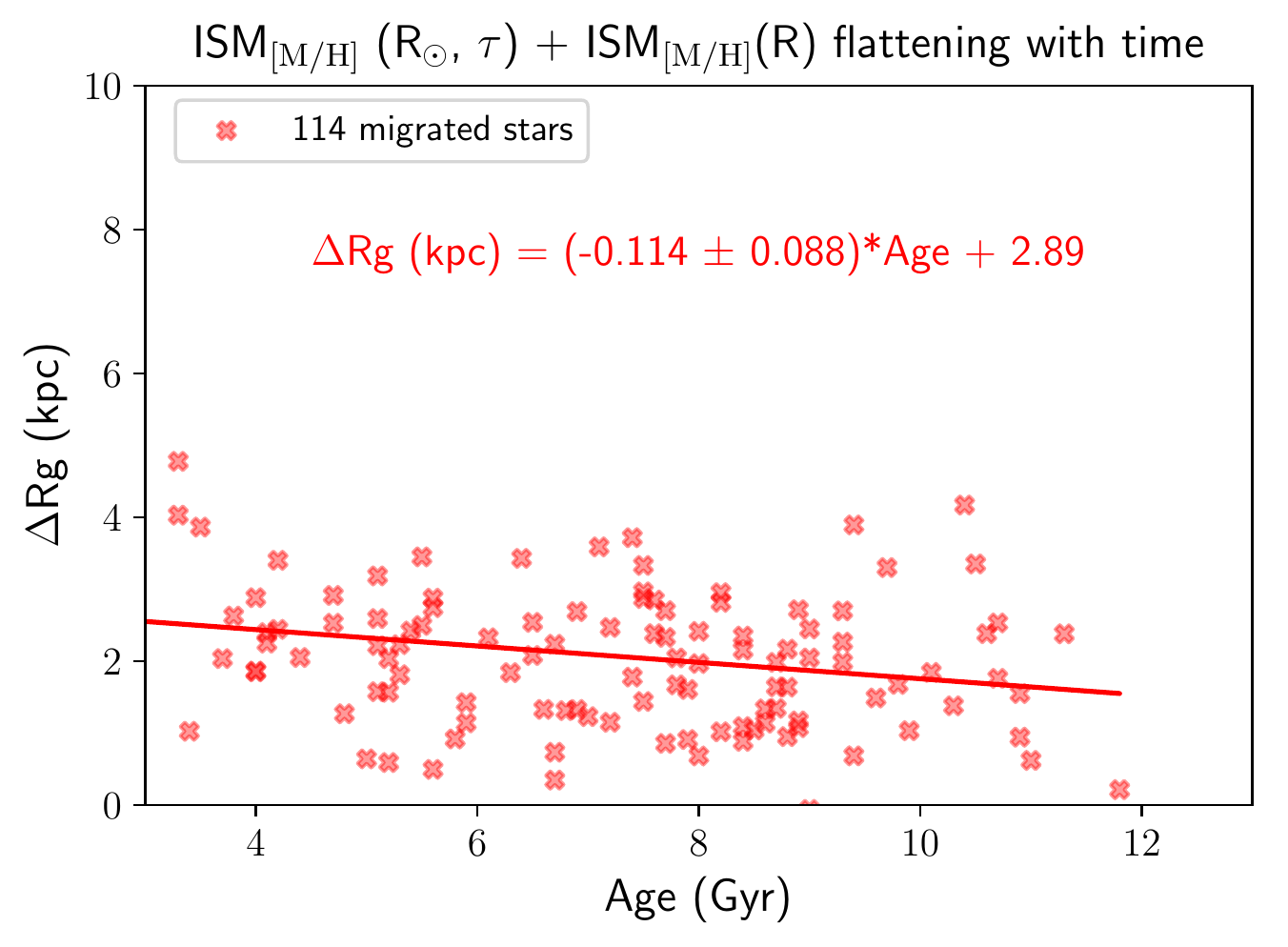}
\caption{Estimated covered distance $\Delta$R$\rm _g$ (R$\rm _g$ - R$_{\rm birth}$) vs. stellar age for the selected churned SMR stars, applying a different zero point (ISM$_{\rm [M/H]}$(R$_\odot$,$\tau$)), and a different ISM metallicity gradient as a function of the stellar age (linearly flattening with time from -0.15 to -0.07 dex~kpc$^{-1}$ between 11.8 Gyr and 3.3 Gyr, respectively) in the R$_{\rm birth}$ estimate.}
\label{Fig:migrated_distance_flattening}
\end{figure}

Figure \ref{Fig:MIGRATED_candidates} shows the orbital eccentricities as a function of [M/H] for the metal-rich disc sample. The solid lines correspond to the required eccentricity (see Equation (\ref{eq:ecc})) for different values of ISM radial metallicity gradients: -0.10~dex~kpc$^{-1}$ (black), \break -0.07~dex~kpc$^{-1}$  \citep[our measured gradient for young stars in Table~\ref{table:gradient}; see also][red]{minchev2018}, -0.04~dex~kpc$^{-1}$ (orange), and -0.06~dex~kpc$^{-1}$ \citep[Cepheids analysis from][green]{genovali2014}. For the three first cases, we assumed ISM$_{\rm [M/H]}$(R$_\odot$)~=~0.0 to estimate R$_{\rm birth}$ from the stellar metallicity. However, \citet{genovali2014} have their own zero point, defined as: [Fe/H]~=~-0.06 * R$_{\rm g}$ + 0.57, with a clear shift in the relation compared to the other ones assumed in this work. The impact of the ISM gradient value and the zero-point assumption on the derived R$_{\rm birth}$, and therefore on the required eccentricity to reach the solar vicinity without the need for churning, is clearly observed. As described in \citet{hayden2020}, given the measured [M/H] and eccentricity, stars lying to the left are able to reach the solar neighbourhood through blurring, while the stars to the right of the line are possible candidates to have migrated through churning. This is the case for most of the SMR stars (70\% of the SMR~stars lie below the line that corresponds to the Cepheids analysis); they are therefore likely to have been brought to the solar neighbourhood by churning, which is in close agreement with previous studies \citep[e.g.][]{georges2015b, wojno2016}. However, it is worth noting that the observed metallicity distribution function in Fig.~\ref{Fig:hist_ALL} peaks around 0.2~dex, which is higher than previous reported solar vicinity MDFs \citep[see e.g.][]{fuhrmann2017}. A possible ignored bias towards more metal-rich objects in the sample selection could be pulling the percentage of possible migrators to higher values. Among the entire distribution, our churned candidates with [M/H]~$\textgreater$~+0.1 comprise around 17~\% of the sample. If we constrain the number of migrators to only stars with [M/H]~$\textgreater$~+0.25, the global percentage decreases to 8~\% of the sample. \par


In particular, we analysed the possible age trends in the radial changes ($\Delta$R$\rm _g$ = R$\rm _g$ - R$_{\rm birth}$) associated to churning. For each assumed ISM abundance gradient separately, Fig. \ref{Fig:migrated_distance} shows the $\Delta$R$\rm _g$ as a function of stellar age for the churned candidates (see Fig.~\ref{Fig:MIGRATED_candidates}). The migrated SMR stars cover a wide range of ages from $\sim$~4 to 12~Gyr. As a consequence of the R$_{\rm birth}$ estimation proxy (zero point fixed at ISM$_{\rm [M/H]}$(R$_\odot$) = 0.0 dex), the flatter the applied ISM gradient, the larger the $\Delta$R$\rm _g$. The respective slopes and errors were derived by the Theil-Sen linear regression estimator. We observe that the general trend is slightly negative with stellar age, becoming steeper as the assumed ISM gradient flattens (see right-hand panel). In addition, we obtain different $\Delta$R$\rm _g$ values based on the zero-point expression from the  analysis of \citet{genovali2014}, but their study also reveals a shallow negative gradient with stellar age. As a consequence, the zero-point assumption has a direct effect on the value derived for migration distance, but not on the observed general trend with age. For instance, if we assume a higher zero point (e.g. ISM$_{\rm [M/H]}$(R$_\odot$)~=~0.1~dex), the radial migration estimate decreases by almost half for a star with [M/H]~=~+~0.25. \par 

Additionally, the application of a fixed zero point for every star may introduce a bias in the $\Delta$R$\rm _g$ estimate, in particular for old stars, because ISM$_{\rm [M/H]}$(R$_\odot$) is expected to have been chemically enriched with time to reach the present-day value used here. To explore the dependence on the zero-point assumption, we applied a more appropriate value for each star according to its age. For this purpose, we selected the observed stars in our sample that lie on the solar annulus (7.5~$\leq$~R$\rm _g$~<~8.5 kpc), and calculated the average metallicity ([M/H]~=~0.02, -0.03, -0.05, and -0.15 dex) at different age bins ([2-6], [6-8], [8-10], and [10-12] Gyr). We then assumed these values as a proxy of the ISM$_{\rm [M/H]}$(R$_\odot$) evolution with time. The adopted trend is consistent with the ISM enrichment of the solar neighbourhood found by Galactic chemical evolution models \citep[e.g.][]{hou2000, schonrich2017, minchev2018}, although an accurate ISM$_{\rm [M/H]}$(R$_\odot$,$\tau$) estimate is outside the scope of this paper. Figure \ref{Fig:migrated_distance_zeropoint} shows the observed trend by applying the estimated zero point as a function of the SMR stellar age. For each analysed ISM abundance gradient, we obtained significantly higher $\Delta$R$\rm _g$ values for older SMR stars up to inverting to a positive trend with stellar age. We stress that the selected stars in the solar annulus are likely to be originally born far from their present location. Therefore, a more accurate ISM$_{\rm [M/H]}$(R$_\odot$, $\tau$) estimate might present lower values than the ones shown here, and the impact on the $\Delta$R$\rm _g$ value could be greater than that suggested by our analysis. \par

Moreover, we explored the influence of a time evolution of the ISM gradient combined with an evolving ISM$_{\rm [M/H]}$(R$_\odot$) zero point  on the R$_{\rm birth}$ estimate. 
For this purpose, we decided to apply a simple toy model based on the assumptions of \citet{minchev2018}: a thin disc formed with an initial metallicity gradient  of $\sim$~-0.15 dex~kpc$^{-1}$, flattening with time to a present-day ISM profile of  $\sim$ -0.07 dex~kpc$^{-1}$ (which is equal to our measured gradient for young stars, shown in Table \ref{table:gradient}). 
Our toy model simply assigns these limit ISM gradient values to the youngest (3.3 Gyr) and the oldest (11.8 Gyr) SMR stars in our selected sample. A linear interpolation was then performed to estimate the corresponding ISM metallicity gradient as a function of each SMR stellar age. Figure \ref{Fig:migrated_distance_flattening} shows the resulting trend of $\Delta$R$\rm _g$ with stellar age, which has now changed sign compared to the one derived in Fig. \ref{Fig:migrated_distance_zeropoint}, but is similar to the observed trend in our first simple approach shown in Fig. \ref{Fig:migrated_distance}. \par 

We would like to highlight the fact that the aim of this analysis is to visualise the strong dependence on the assumptions of the observed radial migration trend with age. An accurate ISM$_{\rm [M/H]}$(R) determination at any epoch is outside the scope of this paper. For that reason, we assumed a linear one-slope gradient for simplicity to estimate the birth radius of the stars for each analysed case. A more accurate metallicity gradient \citep[e.g. flattens at R $\textless$ 6~kpc as found by][]{hayden2014, haywood2019} would probably provide more accurate values \citep[i.e. with no need to go further than 2-3~kpc to find the most metal-rich stars, as shown by the APOGEE results from][]{hayden2014,hayden2015}, but the dependence
of the general trends on the  assumptions of the  illustrated model  would be similar to those reported here. \par

In conclusion, our local analysis hints towards a clear, although not necessary predominant, presence of radially migrated stars in the Galactic disc via churning, in agreement with the findings of \citet{georges2015b}. The more realistic scenario illustrated in Fig.~\ref{Fig:migrated_distance_flattening} (assuming a time evolution of the ISM gradient and the ISM$_{\rm [M/H]}$(R$_\odot$) zero point) shows an average $\textless$~$\Delta$R$\rm _g$~$\textgreater$~=~2.03~$\pm$~0.95~kpc for a broad range of ages  (3.3~$\leq$~$\tau$~$\leq$~11.8~Gyr). In the same way, we estimate an average $\textless$~$\Delta$R$\rm _g$~$\textgreater$~=~2.06~$\pm$~0.84~kpc in Fig.~\ref{Fig:migrated_distance} (left-most panel) for the case of an ISM gradient similar to the measured one in this work ($\sim$~-0.10 dex~kpc$^{-1}$; see Table~\ref{table:gradient}), which is compatible with the measured gradient by \citet{hayden2014} using APOGEE data (shown in Table~\ref{table:gradient3}). This behaviour suggests that an important fraction of stars in the Galactic disc could have also been sensitive to radial changes associated to churning, favouring a scenario where metal-rich stars may come from 2-3~kpc from the Sun. Unfortunately, from an observational point of view, only high-metallicity stars in the solar vicinity can realistically be assumed to be radial migrators. We again mention how complicated it is to accurately measure the radial migration efficiency, and how the interpretation of the reported signatures strongly relies on the model assumptions.

\section{Age--abundance relations} \label{results_2}

In this section, we analyse the Galactic disc as a whole, without considering a thin--thick disc dichotomy, to allow a general, parametrised view of the disc evolution with time.  

\subsection{[Mg/Fe] abundance as a chemical clock} \label{MgvsAGE}

\begin{figure}
\centering
\includegraphics[height=70mm, width=0.48\textwidth]{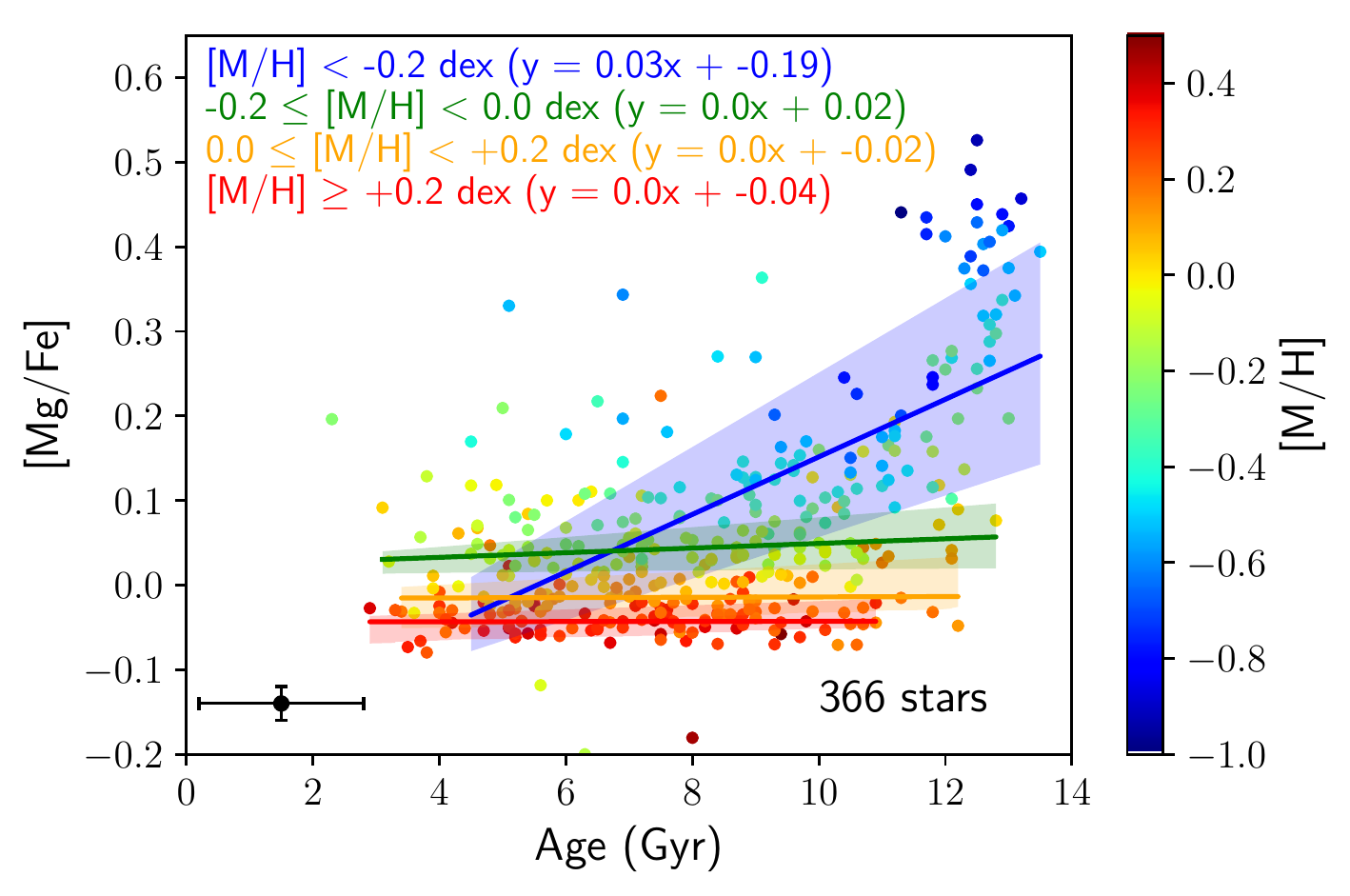}
\caption{[Mg/Fe] as a function of age for the working sample. The coloured lines correspond to the Theil-Sen linear regression over different metallicity ranges, from metal-poor to metal-rich (from top to bottom). The respective shaded areas span the lower to upper bound of the 95\% confidence interval of the fit. The mean estimated uncertainties for each star value are shown in the bottom-left corner.}
\label{Fig:MgAge_M}
\end{figure}

Figure \ref{Fig:MgAge_M} shows the [Mg/Fe] versus age relation colour-coded according to stellar metallicity. 
The figure clearly shows 
a significant spread in stellar age at any given [Mg/Fe] value, particularly for [Mg/Fe] lower than 0.2~dex. The covered age range at a fixed [Mg/Fe] is generally wider in comparison with the \citet{hayden2017} analysis of the same sample of stars (see their Fig.~2) using Gaia DR1 data and abundances from \citet{sarunas2017}. 
Furthermore, for stars younger than about 11 Gyr, we observe a larger dispersion in the [Mg/Fe] abundance ($\sigma_{\rm [Mg/Fe]}\sim$ 0.1 dex at a given age) than that suggest by previous studies \citep[e.g.][]{delgadomena2019, nissen2017, nissen2020}. This dispersion is correlated with the stellar metallicity and is much more apparent thanks to the unveiled slope of [Mg/Fe] with [M/H] in \citet{SantosPeral2020} for metal-rich stars. Indeed, in close agreement with \citet{haywood2013}, the lower envelope is occupied by metal-rich stars, while the upper envelope is occupied mainly by more metal-poor stars. \par

\begin{figure*}
\centering
\includegraphics[height=120mm, width=\textwidth]{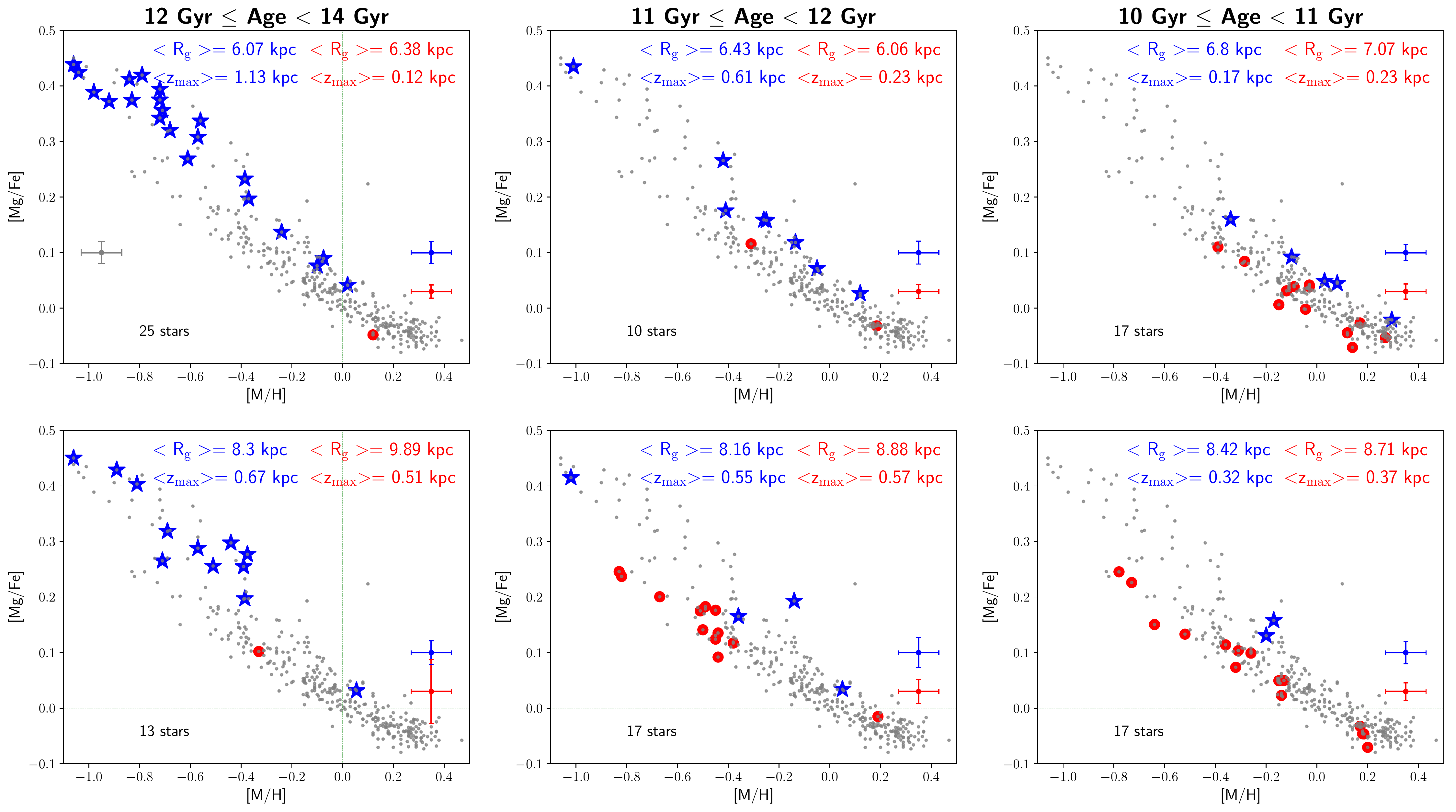}
\caption{Distribution of the selected sample in the ([M/H], [Mg/Fe]) plane at different ages and locations in the Galaxy (R$\rm _g$). Two chemical sequences appear for ages younger than $\sim$ 11-12~Gyr, corresponding to the classical thick (blue stars) and thin (red circles) disc components. Each panel corresponds to a bin in the (age, R$\rm _g$) space, 
dividing the Galactic disc into two regions: inner (R$\rm _g$ $\leq$ 7.5 kpc; top row) and outer (R$\rm _g$ > 7.5 kpc; bottom row). The blue and red crosses in the bottom-right corner of each panel represent the mean estimated errors in [Mg/Fe] and [M/H] for the thick and thin disc population, respectively, at that particular radius and time. The average R$\rm _g$ and z$_{\rm max}$ for each population are given at the top of each panel. The whole working sample is shown by the dotted grey points.}
\label{Fig:MgvsM_old}
\end{figure*}

\begin{figure*}
\centering
\includegraphics[height=120mm, width=\textwidth]{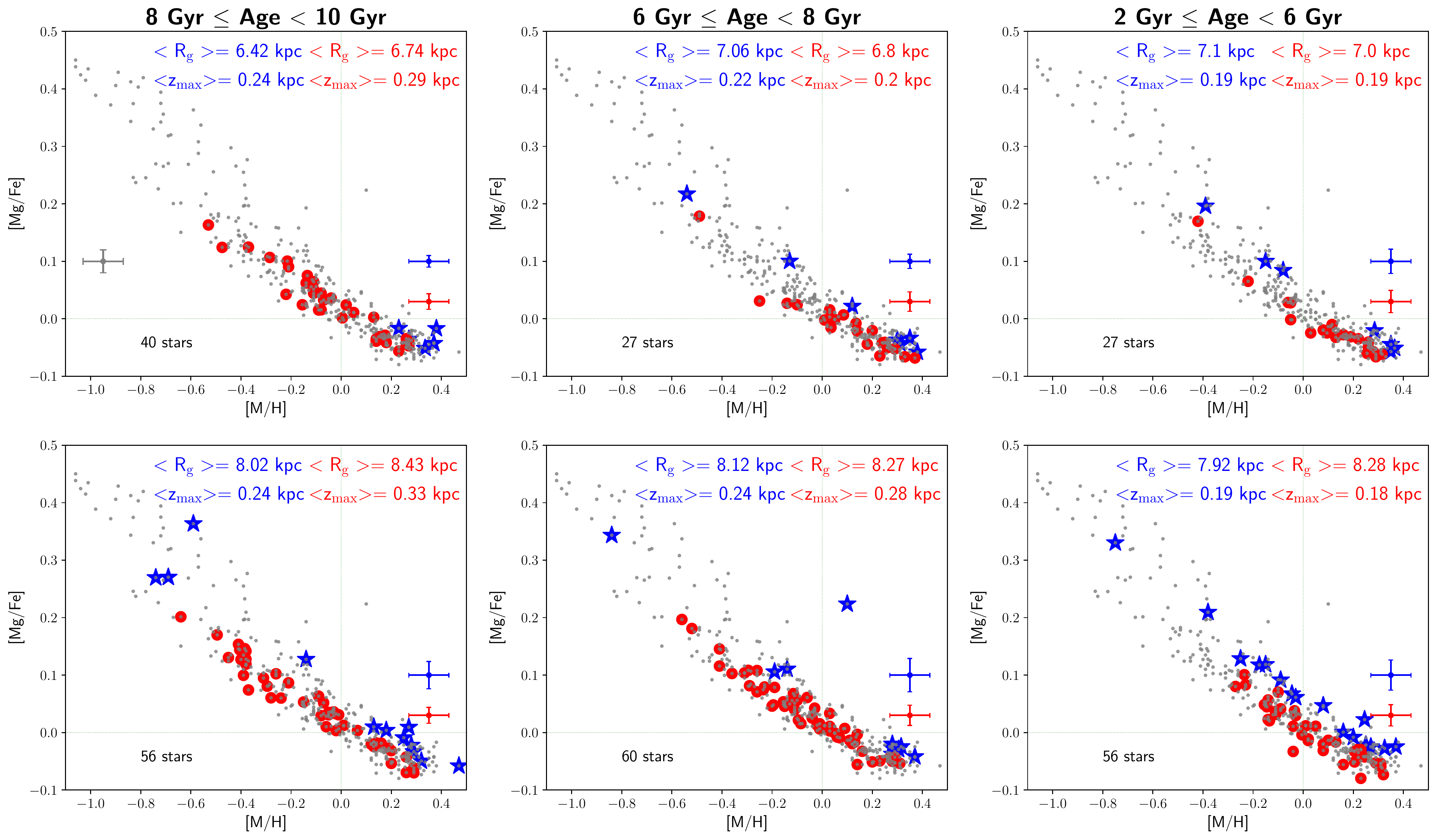}
\caption{Same as Fig. \ref{Fig:MgvsM_old} but for the younger age intervals.}
\label{Fig:MgvsM_young}
\end{figure*}

As a follow up, we studied the [Mg/Fe]--age trends in different metallicity bins. 
For [M/H]~$\geq$~-0.2, we find stars with ages from $\sim$ 3 to 12 Gyr, describing a flat trend in the [Mg/Fe]--age plane,  without a change of the slope in the different metallicity bins. This observed pattern could only be explained by chemical evolution models if an important co-existence of different stellar populations in the solar neighbourhood is assumed, with different enrichment histories and birth origins in the Galactic disc \citep[as previously suggested by several studies, e.g.][]{sellwood2002, nordstrom2004, fuhrmann2011, boeche2013b, georges2015a, wojno2016}. \par 

For more metal-poor stars ([M/H]~<~-0.2~dex), we find a linear correlation of [Mg/Fe] with age, showing a consistent positive slope value with the results of \citet{delgadomena2019} ($\sim$0.02~dex~Gyr$^{-1}$, see their Fig.~7 for a  sample of dwarf thin disc stars in the solar neighbourhood), and \citet{ness2019} ($\sim$0.03~dex~Gyr$^{-1}$; see their Fig.~7 for a sample of red clump stars across a wide range of Galactocentric distances). However, these works do not find a clear trend of [Mg/Fe] abundances with metallicity in the metal-rich regime such as the one we report for our sample. This slope is also very consistent with the one derived for sulphur by \citet[][]{perdigon2020}. \par 

Finally, we note the presence of young, metal-poor, and high-[Mg/Fe] stars, which has also reported in the literature \citep[e.g.][]{haywood2013, martig2015,chiappini2015, fuhrmann2017_BSS1, fuhrmann2017_BSS2, silvaaguirre2018,delgadomena2019, ciuca2020}. Such stars have been suggested to be radially migrated candidates expelled outwards by the Galactic bar \citep{chiappini2015}, or blue straggler stars produced by mass transfer in binary systems \citep{jofre2016, wyse2020}, which may lead to underestimation of their age.

\subsection{Temporal evolution in the [Mg/Fe]-[M/H] plane} \label{MgvsM_time}

Figures \ref{Fig:MgvsM_old} and \ref{Fig:MgvsM_young} illustrate the [Mg/Fe] abundance ratios relative to [M/H] for our sample of stars, in different age intervals and Galactic disc locations (top row: inner Galactic disc R$\rm _g$~$\leq$~7.5~kpc; bottom row: outer disc R$\rm _g$~>~7.5 kpc). The bin selection was optimised to allow a significant statistical sample of stars at different radii. \par 

Figure \ref{Fig:MgvsM_old} shows that 
the oldest stellar population  ($\tau$~$\geq$~12~Gyr) at every radius is located in what has been classically called the thick disc, that is, an $\alpha$-enhanced population (see Sect.~\ref{disc_distinction}). 
During this early epoch, we observe rapid chemical enrichment, reaching solar metallicities. In addition, the stars are more centrally concentrated (see Sect.~\ref{metalrich_distinction}). Interestingly, 11-12 Gyr~ago, a second chemical sequence appears in the outer regions of the Galactic disc, populating the metal-poor low-[Mg/Fe] tail and starting at [M/H]~$\sim$~-0.8~dex and [Mg/Fe]~$\sim$~0.2~dex. This corresponds to the population that has classically been referred to as the thin disc or low-$\alpha$ population, highlighted in the figure by red points. Based on the radial extension of the thick disc sample (see Figs.~\ref{Fig:hist_DIST_rich} and \ref{Fig:hist_DIST_poor} in Sect.~\ref{metalrich_distinction}), the average R$_{\rm g}$ of this second chemical sequence seems to be significantly larger than the Galactic disc extension at that time (R$_{\rm g}$~<~8.5~kpc), reaching the outer parts up to R$_{\rm g}$~$\sim$~11~kpc. Furthermore, those stars are shown to be significantly more metal-poor ([M/H]~$\lesssim$~-0.4~dex) with respect to the coexisting stellar population in the inner parts of the disc (top row). They also show lower [Mg/Fe] abundances than the older disc population in the outer parts (lower left panel), although presenting a similar metallicity distribution. This implies a chemical  discontinuity in the disc around 11~Gyr~ago, suggesting that the new sequence might have followed a different chemical evolution pathway from that followed by the previously formed component, possibly triggered by accretion of metal-poor external gas \citep[e.g.][]{grisoni2017, noguchi2018, spitoni2019, buck2020, palla2020}. However, as a direct comparison with models and simulations is missing in the interpretation of our observational results, we are not able to discard other possible scenarios (e.g. separate evolution of the outer disc). \par


Figure \ref{Fig:MgvsM_young} illustrates the chemical evolution in the last 10~Gyr. During this long period of time, the Galactic disc seems to have experienced a slower and more continuous chemical evolution towards more metal-rich and lower [Mg/Fe] regimes than the more primitive epochs described in Fig. \ref{Fig:MgvsM_old}. In addition, the average z$_{\rm max}$ seems to decrease with age, regardless of the chosen chemical sequence and space location, which could be a signature of a disc settling with time. However, a larger statistical sample would be needed to justify this assumption. Finally, the difference in z$_{\rm max}$ between both chemical sequences does not allow any relevant conclusion, given the low number of stars and the same followed z$_{\rm max}$ distribution (the present similarities do not support two distinct disc populations; see Sect.~\ref{metalrich_distinction}).

\subsection{Trends with stellar age: relation to radius}

\begin{figure*}
\centering
\includegraphics[height=70mm, width=0.48\textwidth]{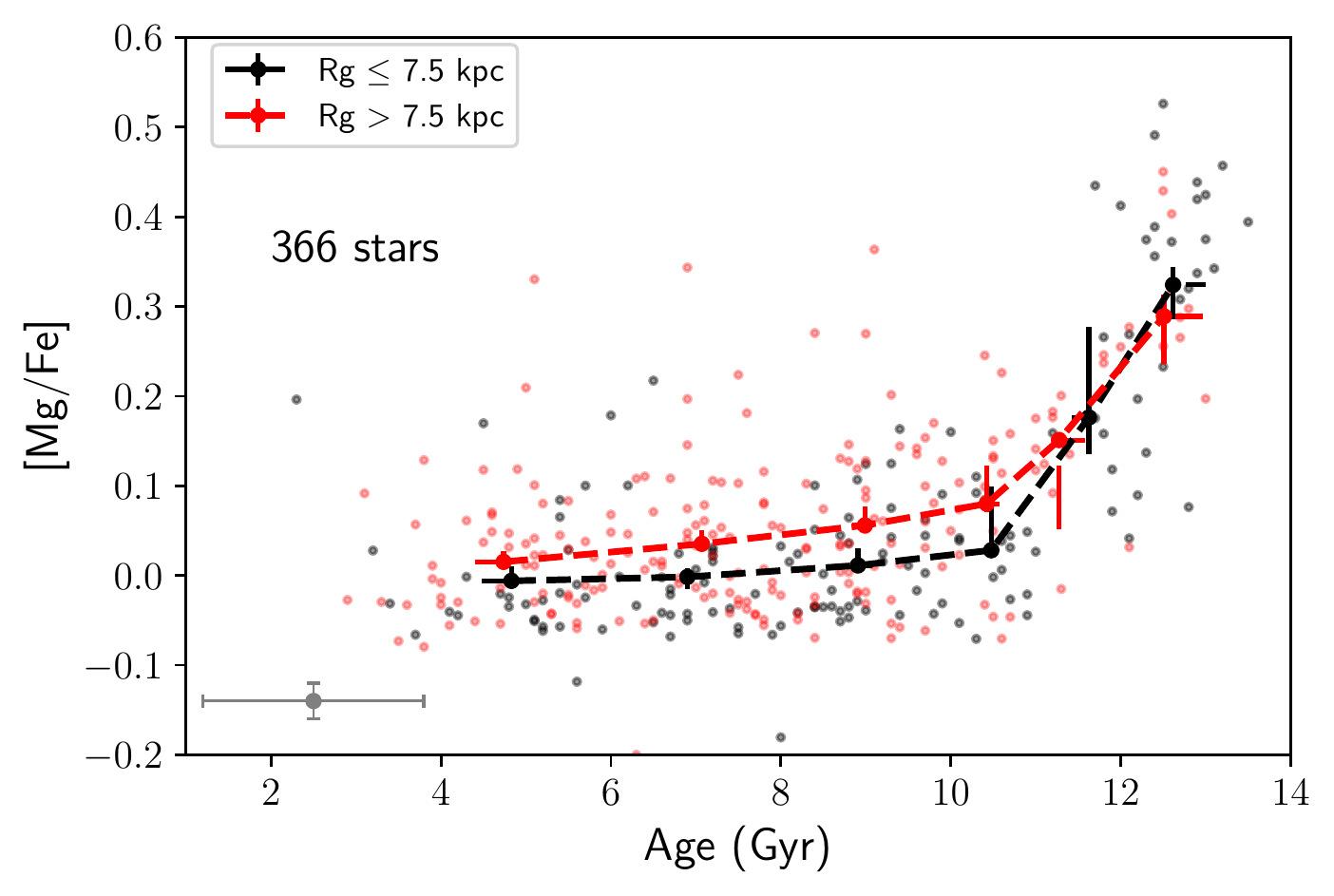}
\includegraphics[height=70mm, width=0.5\textwidth]{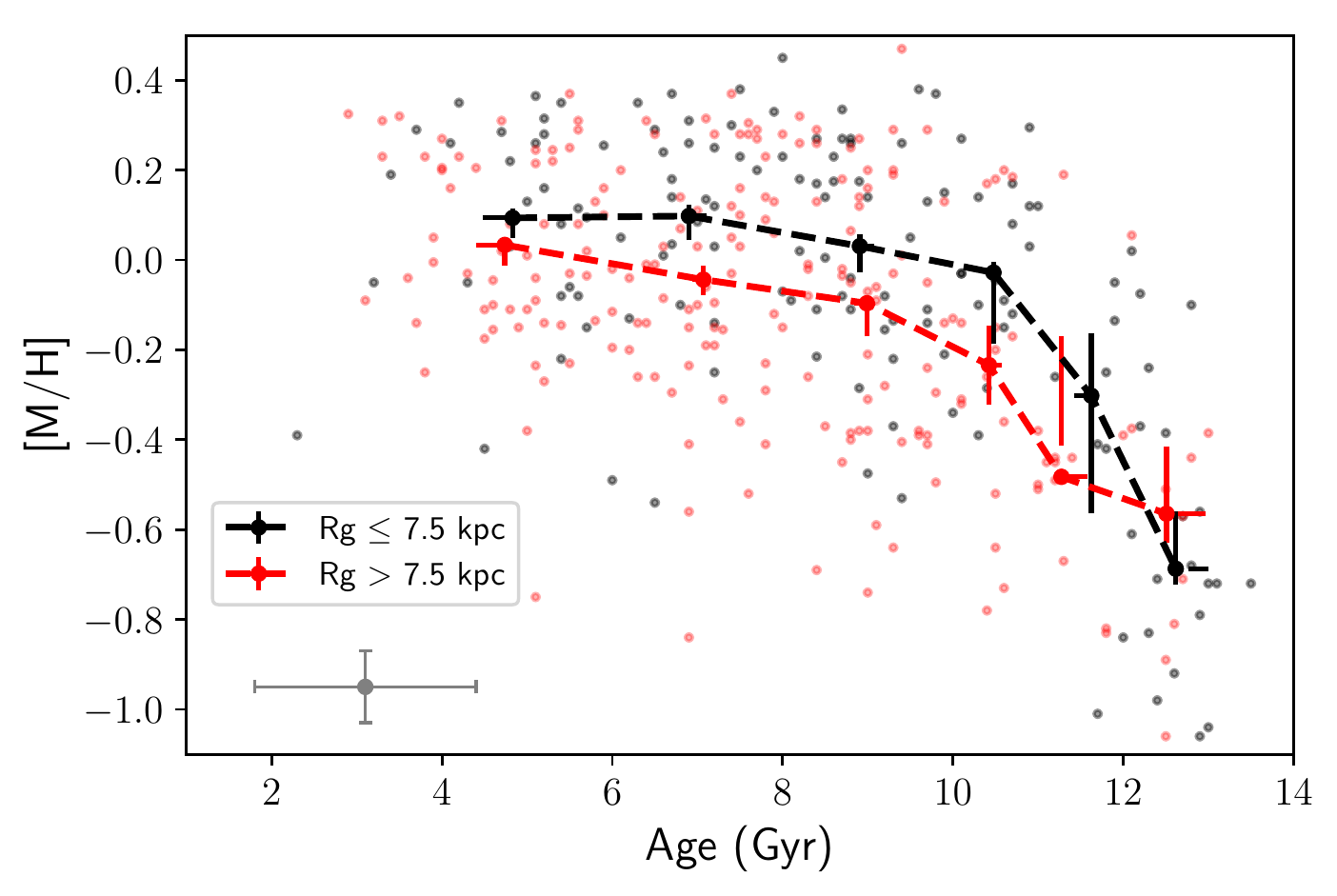}
\caption{Stellar distribution in the [Mg/Fe] vs. age (left panel) and the [M/H] vs. age relation (right panel) for the inner (R$\rm _g$ $\leq$ 7.5 kpc) and the outer (R$\rm _g$ > 7.5 kpc) regions of the Galactic disc. Their respective trends are indicated by the average abundance value per age bin ([2-6], [6-8], [8-10], [10-11], [11-12] and [12-14] Gyr), together with the error bars in [Mg/Fe], [M/H], and age. 
The overall mean estimated errors are shown in the bottom-left corner.}
\label{Fig:AMR_Rg}
\end{figure*}

Figure \ref{Fig:AMR_Rg} represents the [Mg/Fe] versus age (left panel) and the age--metallicity relation (AMR; right panel), in two bins of guiding centre radius: the inner regions (R$\rm _g$ $\leq$ 7.5 kpc) and the outer regions (R$\rm _g$ > 7.5 kpc). The points correspond to the average abundance estimate for each bin in the (age, Rg) space, shown as individual panels in Figs. \ref{Fig:MgvsM_old} and \ref{Fig:MgvsM_young}. \par

In order to propagate the uncertainties on [M/H], [Mg/Fe], and age for each star, we performed 1000 Gaussian Monte Carlo realisations within the corresponding error bars of each object. In addition, to include the effect of possible statistical fluctuations in the sampled population, we randomly selected 80 \% of the stars for each Monte Carlo realisation, and estimated the average values in age, [M/H], and [Mg/Fe] for the different bins in age. Finally, the plotted errors correspond to the 16th and 84th percentile values ($\pm$ 1$\sigma$) of the resulting distributions of the average age, [M/H], and [Mg/Fe] abundances. We did not consider the uncertainties in the R$\rm _g$ estimation because they are negligible (smaller than 0.2~kpc, see Sect.~\ref{orbits}). \par

 Figure \ref{Fig:AMR_Rg} shows a rapid chemical enrichment at early epochs (10-14 Gyr), with a sharp increase in metallicity and decrease in [Mg/Fe] to solar abundance values. The inner and outer regions describe similar curves, which may indicate that the two samples were drawn from a similar set of stars. We reiterate that  
the outer disc population does not extend farther than $\sim$ 8.5~kpc before $\sim$11-12 Gyr. \par  

Subsequently, the appearance of the second Galactic disc sequence separates the average values of the inner and outer disc bins. In particular, the outer bin seems to move away from the initial tracks, with a flatter behaviour in [M/H] versus age around 11~Gyr.
From that epoch, we find two approximately parallel patterns for the inner and outer disc, flattening in the same way with time for the last 10~Gyr of evolution. The measured [M/H] ([Mg/Fe]) abundance is shifted towards lower (higher) values for the outer disc at any age, which corresponds to the measured negative (positive) radial gradient in Sect.~\ref{gradients}. The inside-out scenario for the disc build-up \citep{matteucci1989, chiappini2001}, that is, larger timescales for larger radii, which reproduces the observed radial gradients in the disc and the higher surface gas density for star formation in the inner regions, agrees with our observed result. Although our limited sample does not allow us to interpret the disc evolution at different galactocentric distances or smaller bins in R$_{\rm g}$, it is worth noting that the similar observed slopes of the [Mg/Fe] and [M/H] trends with age may be the signature of a similar chemical evolution at different radii for ages younger than 10~Gyr. \par 

Finally, the shift in the abundance space is particularly visible in the outer disc bin and the age interval of 11-12~Gyr (right panel of Fig. \ref{Fig:AMR_Rg}) because of the appearance of the thin disc sequence, leading to a misalignment between the average abundance value and the error bar for this bin. We find that the Monte Carlo realisations introduced a bias in this particular bin due to the random contamination of younger and older stars in the propagation of the stellar age uncertainties. Compared to the original sample shown in Fig.~\ref{Fig:MgvsM_old}, the inclusion of stars younger than 11~Gyr shifts the metallicity distribution to significantly higher values, while stars older than 12~Gyr do not compensate this bias due to their similar metallicity composition ([M/H] $\lesssim$ -0.4~dex).

\section{Discussion on the formation and evolution of the Galactic disc} \label{discussion}



Different formation and evolution scenarios for the Milky Way disc have been proposed in order to interpret or reproduce the present-day observed chemodynamical trends in the disc. In particular, the chemical bimodality identified in the ([$\alpha$/Fe], [M/H]) plane as the thin and thick disc components, and the observed radial abundance gradients, encode information on how the discs have evolved. The main ingredients of disc evolution (mass accretion and radial migration) are examined below in light of our results. \par 

Our results are purely observational and the described interpretation hereafter is based on a revision of literature studies and does not take into account or make direct comparisons with models or simulations. We are aware of this limitation in our interpretation, and a detailed comparison between our data and model results will be performed in a future work, where we hope to draw more robust conclusions.

\subsection{Importance of internal evolutionary processes}

One of the main model assumptions influencing disc evolution is the importance of the Milky Way environment. It is a known fact that the Milky Way is not evolving in isolation, but the role of internal secular processes in shaping the $\alpha$-elements bimodality is difficult to evaluate. To tackle this point, some studies have tested the reproducibility of the disc chemodynamical features involving minor quantities of mass accretion and mass loss. In particular, the importance of the following internal processes has been explored: 
(i) radial flows of gas and radial migration of stars playing a significant role in the stellar distribution \citep[e.g.][]{schonrich2009, bovy2016, mackereth2017, sharma2020}, and (ii) the inflow of metal-poor left-over gas from the outskirts of the disc \citep{haywood2019, khoperskov2020, katz2021} triggered by the formation of the bar and the establishment of the outer Lindblad resonance. For instance, \citet{khoperskov2020} developed self-consistent chemodynamical simulations of Milky Way-type galaxies that reproduce the observed $\alpha$-bimodality on a long time-scale, after chemical enrichment of the surrounding halo by stellar feedback-driven outflows during the fast thick disc formation.


Our results present a clear [Mg/Fe] chemical distinction between both Galactic disc sequences in the metallicity regime [M/H] $\leq$ -0.3 dex, and a general decreasing trend even at supersolar metallicities ([M/H] > 0). 
We find that the oldest observed population ($\tau$~>~12~Gyr) constitutes the majority of the more metal-poor high-[Mg/Fe] stars, in agreement with the results of previous observational studies of the local disc \citep[e.g.][]{fuhrmann2011, haywood2013, hayden2017}, where a temporal transition between the high- and low-[Mg/Fe] populations  was reported as being due to two different star formation epochs. 
Furthermore, we identified the presence of the metal-poor low-[Mg/Fe] stellar population in the outer regions of the Galactic disc (up to R$_{\rm g}$~$\sim$~11~kpc) around 10-12 Gyr ago \citep[old metal-poor low-alpha disc stars have also been found before; e.g.][]{haywood2013, delgadomena2019, buder2019}. These stars appear significantly more metal-poor ([M/H] $\lesssim$ -0.4 dex) than the coexisting stellar population in the inner parts of the disc. For the next 10 Gyr of evolution up to the present day, we do not observe significant differences in the chemical enrichment of the inner and outer regions, both presenting similar slopes with respect to the stellar age. \par

We evaluated our results considering the above-mentioned works that test dominant internal evolution processes. 
First, regarding radial migration, our estimates in Sect.~\ref{migration} support an important efficiency of the churning process, although the actual amplitude of the induced radial changes remains uncertain. Second, the scenario whereby the metal-poor thin disc stars originate from an inflow of gas from the outskirts of the disc \citep{haywood2013, haywood2019} could be challenged by the similar continuous behaviour of the chemical tracks with radius, which do not show enrichment discontinuities between the internal and the external regions. Indeed, as mentioned above, after the appearance of the low-[Mg/Fe] sequence around 10-12~Gyr ago, our results show similar chemical evolution trends ([Mg/Fe] vs. age, and AMR) for the inner and the outer Galactic disc; they describe separated patterns, where the measured metallicity is always higher in the inner regions at any age, describing a negative radial gradient. Although our local sample does not allow us to accurately study the disc at different Galactocentric distances, as in the recent works by \citet{haywood2019} and \citet{katz2021}, the observed behaviour may indicate that the stellar Galactic disc has likely followed the same chemical evolution path but at a different radius, without any apparent radial discontinuity. This does not preclude a separate evolution of the external disc before this secondary phase 10-12~Gyr ago, but it would imply a change in the radial dependence of the chemical evolution, which would need to be explained. Further analysis with a larger disc stellar sample, beyond the solar vicinity region, is requied in order to draw more robust conclusions.\par


\subsection{Mass accretion}

Recent analyses based on \emph{Gaia} DR2 data have revealed that the Galaxy is perturbed from equilibrium, finding evidence that the Galactic disc can be strongly perturbed by external interactions with satellite galaxies \citep[][latter work with current EDR3 data]{antoja2018, laporte2019c, antoja2020}, affecting its dynamics and star formation history, although there is no general consensus on the interpretation \citep[see also][]{khoperskov2019, bennett2020}. 
In line with these findings, \citet{helmi2018} and \citet{belokurov2018} \citep[see also][]{haywood2018} found clear signatures of a major accretion event by a massive satellite approximately 10 Gyr ago, the so-called \emph{Gaia}-Enceladus/Sausage, which seems to be considered as a key piece of the puzzle. In this context, models invoking a steady Galactic star formation history seem to neglect a very important ingredient of Milky Way history: mass accretion events. 
\par

A variety of disc evolution scenarios allow an infall of mass from the Milky Way environment. This includes chemical evolution models predicting infall of gas on long timescales \citep{chiappini1997, spitoni2020}, and cosmological $\Lambda$CDM simulations \citep{buck2020, agertz2020} describing mass accretion through mergers of satellites and gas.
In these models, the disc formation goes through two separated star formation epochs driven by two main gas-accretion episodes of extragalactic pristine gas, triggering two different chemical evolution pathways \citep[e.g.][]{chiappini1997, grisoni2017, grisoni2018, noguchi2018, spitoni2019, buck2020, palla2020, montalban2020}.



Recent cosmological hydrodynamical simulations developed by  \citet{buck2020} reproduce the bimodal sequences in the [$\alpha$/Fe] versus [Fe/H] plane as a consequence of a gas-rich merger during the evolution of the  Galaxy. In agreement with our results, their models recreate a formation scenario where the high-$\alpha$ sequence evolves first in the early galaxy, pre-enriching the ISM to high metallicities before the appearance of the low-$\alpha$ sequence. The second low-$\alpha$ sequence is formed after the gas-rich merger, which provides metal-poor gas, diluting the ISM metallicity. In all simulations, these latter authors observed an old (younger), more radially concentrated (extended) high (low)-$\alpha$ disc, compatible with our results. Additionally, the recent work by \citet{agertz2020}, using cosmological zoom-in simulations of a Milky Way-mass disc galaxy, found a similar connection between a last major merger at z~$\sim$~1.5 and the formation of an outer, metal-poor, low-$\alpha$ gas disc around the inner, in-situ metal-rich galaxy containing the old high-$\alpha$ stars. 

There is an apparent overlap in time and metallicity 
between the reported formation of the outer metal-poor low-[Mg/Fe] disc in our results and the signature of a major merger by a massive satellite found in the literature \citep{helmi2018, belokurov2018, myeong2018, DiMatteo2019, gallart2019, emma2019, belokurov2020, feuillet2020, naidu2020, georges2020}. Therefore, our results seem to be consistent with the arrival of a pristine gas-rich merger after the formation of the pre-existing old high-[Mg/Fe] population, which may have been diluted at different Galactic radii, leading to the formation of the second [Mg/Fe] sequence on longer timescales. In other words, we find that the Galaxy could have already formed a significant population of thick-disc stars before the infall of the satellite galaxy \emph{Gaia}-Enceladus/Sausage, which could have triggered the formation of the thin-disc sequence more than 10~Gyr ago. \par 

Confirmation of this proposed scenario could be provided by a comparison with chemical evolution models. For instance, the recent work by \citet{palla2020}, who applied new updated two-infall models to reproduce observational data from the APOGEE survey, discuss several cases where the thick and the thin disc are formed by enriched infall episodes including the chemical composition of \emph{Gaia}-Enceladus/Sausage, finding that this satellite merger could have only contributed to the formation of the thick disc. Their best model suggests a delayed two-infall model (gap of around $\sim$ 3.25~Gyr between the formation of the two discs), in agreement with a previous study by \citet[][delay of~$\sim$~4.3~Gyr]{spitoni2019}, with a variable star formation efficiency and the presence of radial gas flows during  evolution of the Galactic disc. Further, detailed analyses are currently being performed  by these latter authors, who are directly comparing their models with our data. The obtained results will be described in a future paper (Palla et al. 2021, in prep.). 

Our interpretation of our observations is that a pre-existing old high-[Mg/Fe] population set the initial chemical conditions in the inner regions (R$\rm _g$ < 8.5 kpc).
As a higher surface density towards the Galactic centre is expected already at early epochs, the pristine accreted gas from \emph{Gaia}-Enceladus/Sausage could have been diluted with a progressively increasing amount of in situ enriched ISM gas towards the disc inner regions. Beyond the limiting radius of the more compact early disc, the infalling gas could have triggered the formation of stars with a chemical composition equal to the accreted one. This scenario would be consistent with the observed chemical similarities between the more metal-rich Gaia-Enceladus/Sausage stars and the metal-poor thin disc population, which are located in neighbouring regions of the [Mg/Fe] versus [M/H] plane \citep[see e.g.][]{helmi2018, myeong2019,feuillet2020}. In addition, this scenario would reproduce the observed chemical radial gradients and the similar observed evolution with age in the inner and outer regions.

\subsection{Radial mixing of stars or gas} 

The interpretation of the abundance gradient evolution with time might be misleading if an undetermined fraction of stars were actually born far from their present location \citep{magrini2009}. Radial stellar migration can decouple the observed evolution of stellar chemical gradients from that of the gradients found in the ISM  \citep[][]{roskar2008, pilkington2012, schonrich2017}.

To account for this uncertainty, our analysis of Sect.~\ref{migration} considers the impact of different zero points and metallicity gradients in the estimates of stellar migration via churning. All the considered radial distributions imply a significant fraction of churned stars in the solar neighbourhood, even considering a slightly super-solar metallicity for the present-day abundance \citep[e.g.][]{genovali2014}. Nevertheless, an important uncertainty remains in the amplitude of the induced radial changes. Indeed, some of the explored scenarios in Figures \ref{Fig:migrated_distance}, \ref{Fig:migrated_distance_zeropoint}, and \ref{Fig:migrated_distance_flattening} imply R$\rm _g$ variations smaller than 1-2~kpc through the entire disc evolution. We can therefore not rule out the possibility that the observed evolution of stellar radial gradients with age is a good first approximation of the actual time-evolution of the ISM. From a dynamical point of view, the possibility  of an intense radial migration inducing only small changes in R$\rm _g$  of the stellar distribution is out of the scope of the present paper. A similar analysis by \citet{feltzing2020} also shows an effective radial migration where some fraction of stars have not moved radially since they formed. It is also worth noting that the effect of radial migration on the [$\alpha$/Fe] versus [M/H] plane deserves further investigation. \par

Finally, radial gas flows could occur in a scenario invoking gas accretion to explain the low-$\alpha$ sequence formation. It is indeed physically plausible that the infalling gas could have had lower angular momentum than the circular motions in the disc, inducing a radial gas inflow towards the inner parts \citep{SpitoniMatteucci2011, schonrich2017}. This would contribute to the dilution of the pre-enriched ISM abundance, shaping the radial gradient 10-12~Gyr ago.

\subsection{Chemical enrichment efficiency and star formation rate} 
At early times (10 $\leq$ $\tau$ < 14 Gyr), the observed high-[Mg/Fe] population in our analysis presents a steep age--metallicity relation, showing an active chemical enrichment, pre-enriching the ISM up to the metal-rich regime before the second low-[Mg/Fe] sequence started to form. These observations are consistent with chemical evolution models predicting an early and fast thick disc formation from a massive gaseous disc ---mainly supported by turbulence--- with a strong star formation rate (SFR) and well-mixed ISM, likely triggered by numerous mergers and accretion events \citep{haywood2013, snaith2014, nidever2014} or due to the rapid collapse of the halo \citep[e.g.][]{prantzos1999, pilkington2012, georges2017}.

In the last 10~Gyr, our analysis shows a flattening with time of the chemical enrichment (c.f. Fig.~\ref{Fig:AMR_Rg}), with a similar slope for the inner and outer regions. This suggests a lower star formation rate with no major radial dependence in the absence of an efficient radial migration. If a radius-dependent SFR were at play, with an increasing SFR with decreasing radius (as expected if the gas density decreases outwards), the  radial migration would likely compensate its effects to mimic a similar chemical enrichment with age at all radii.  Indeed, as the stellar density is higher in the inner Galaxy, there are more stars migrating outwards than inwards. Therefore, in the presence of active radial migration a radius-dependent SFR is compatible with our observations. In this framework, the evolution of radial gradients with time will also be affected, inducing a progressive flattening.

\section{Summary} \label{summary}

We carried out a detailed chemodynamic analysis of a sample of 494 MSTO stars in the solar neighbourhood (d < 300~pc with respect to the Sun) observed at high spectral resolution (R~=~115000) and parametrised within the context of the AMBRE Project. From this sample, we selected 366 stars for which we estimated accurate ages and kinematical and dynamical parameters using the accurate astrometric measurements from the \emph{Gaia} DR2 catalogue. \par

For the analysis, we used the [Mg/Fe] abundances from \citet{SantosPeral2020}. Thanks to careful treatment of the spectral continuum placement for the different stellar types and each particular Mg line separately, we obtained a significant improvement of the precision of abundances  for the metal-rich population ([M/H] > 0~dex), observing a decreasing trend in the [Mg/Fe] abundance even at supersolar metallicites. In contrast, preliminary observational studies of the solar neighbourhood revealed a flattened trend. Therefore, in the present work we studied their impact on the reported chemodynamical features (radial chemical abundance gradients, role of radial migration or `churning', and age--abundance relations), and therefore on the interpretation of the Galactic disc evolution. \par

First of all, we explored the present-day chemical abundance distribution ([Mg/Fe], [M/H]) in the thin disc at different Galactic radii and ages, after applying a chemical definition to avoid possible thick-disc contamination:

\vspace{0.2cm}

\begin{itemize}
    \item \underline{\textbf{Radial chemical trends:}}  We adopted the guiding centre radius (R$\rm _g$) as an estimate of the current stellar Galactocentric radius (R$_{\rm GC}$) to derive the radial abundance gradients. We find a consistent negative gradient of -0.099 $\pm$ 0.031 dex~kpc$^{-1}$ for [M/H] and a slightly positive gradient of +0.023 $\pm$ 0.009 dex~kpc$^{-1}$ for the [Mg/Fe] abundance. The observed steeper [Mg/Fe] gradient than that found in the literature is a major consequence of the \citet{SantosPeral2020} improvement of the metal-rich spectral analysis. In the framework of the time-delay model, this result for the thin disc could be explained by an inside-out formation with a steeper decrease in the star formation efficiency with radius, enhancing the chemical enrichment in the inner regions relative to outer ones. 
    
    \vspace{0.2cm}
    
    \item \underline{\textbf{Stellar migration:}} Under the assumptions of a set of ISM gradients, we estimated the birth radius (R$_{\rm birth}$) of the stars in our sample, given their [M/H], and selected the super-metal-rich stars ([M/H]~$\gtrsim$~+0.1~dex). These stars comprise 17\% of the whole sample (8~\% when we constrain to [M/H]~$\textgreater$~+0.25), and cover a wide range of ages from $\sim$~4~to~12~Gyr. \par
    
    In particular, we analysed the possible age trends in the radial changes ($\Delta$R$\rm _g$ = R$\rm _g$ - R$_{\rm birth}$) associated to stellar migration via churning. First, we fixed the zero point to the present-day value (ISM$_{\rm [M/H]}$(R$_\odot$) = 0.0 dex), observing a slightly negative trend of $\Delta$R$\rm _g$ with age, which becomes steeper as the assumed ISM gradient flattens. Secondly, we applied an approach to estimate a more appropriate ISM$_{\rm [M/H]}$(R$_\odot$) value, which is expected to be chemically enriched with time \citep{hou2000, schonrich2017}. For each analysed ISM abundance gradient, we obtained significantly higher $\Delta$R$\rm _g$ values for old SMR stars up to inverting to a positive trend with stellar age. Finally, we also considered a time evolution of the ISM gradient by applying a simple linear toy model based on an initial metallicity gradient flattening with time \citep{minchev2018}. The resulting slightly negative trend with stellar age is similar to the observed trend in our first simple approach.  \par

    Our analysis showed that regardless of the chosen ISM model, there is a large range of ages for the selected churned stars. This trend suggests that different stars in the Galactic disc could have radially migrated during their lives. However, the radial migration efficiency is very dependent on the adopted ISM gradient \citep{feltzing2020}. 

\end{itemize}

\vspace{0.2cm}

Secondly, in order to shed light on the history of the Milky Way disc, we explored the time evolution of the observed chemodynamical relations. 
We considered the Galactic disc as a whole for the evolution analysis:

\vspace{0.2cm}

\begin{itemize}
    \item \underline{\textbf{[Mg/Fe] abundance as a chemical clock:}} We find a significant spread of stellar ages at any given [Mg/Fe] value, particularly for [Mg/Fe] higher than -0.2~dex, while for more metal-poor stars we find a linear correlation of [Mg/Fe] with age.
    Furthermore, for stars younger than about 11 Gyr, we observe a large dispersion in the [Mg/Fe] abundance ($\sigma_{\rm [Mg/Fe]}\sim$ 0.1 dex at a given age), which is directly correlated with the stellar metallicity. For [M/H]~$\geq$~-0.2~dex, we find stars with ages from $\sim$~3 up to 12 Gyr old, describing an almost flat trend in the [Mg/Fe]--age relation without significant changes of slope among the different metallicity bins. 

    \vspace{0.2cm} 
    
    \item \underline{\textbf{Chemical discontinuity in the [Mg/Fe] versus [M/H] plane:}} We studied the chemical evolution of the [Mg/Fe] abundance relative to [M/H] at different age intervals and Galactic disc locations: inner (R$\rm _g$~$\leq$~7.5~kpc) and outer (R$\rm _g$~>~7.5~kpc). At early epochs (10-14 Gyr), we observed a rapid chemical enrichment, with a sharp increase in metallicity and decrease in [Mg/Fe] to solar abundance values, without apparent differences between the inner and outer regions, 
    pointing towards a chemically well-mixed population \citep{haywood2013,nidever2014, georges2017}. \par
    
    Interestingly, around 10-12 Gyr~ago, a second sequence appears in the outer regions of the Galactic disc, populating the metal-poor low-[Mg/Fe] tail. The average R$_{\rm g}$ of this second sequence is significantly larger than the Galactic disc extension at that epoch (R$_{\rm g}$~<~8.5~kpc), reaching the outer parts up to R$_{\rm g}$~$\sim$~11~kpc. Furthermore, these stars are more metal-poor with respect to the coexisting stellar population in the inner parts, and show lower [Mg/Fe] abundances than the prior disc population in the outer parts, although presenting a similar metallicity distribution. This implies a chemical discontinuity in the disc, also observed in the [Mg/Fe] and [M/H] trends with age, suggesting that the new sequence might have followed a different chemical evolution pathway from that of the previous formed component \citep{grisoni2017, noguchi2018, spitoni2019, buck2020}. It is worth noting that the present result does not support a spatial discontinuity of the chemical evolution in Galactic radius. \par 
    
    From that epoch, 10~Gyr ago, the Galactic disc seems to have experienced a slower and more continuous chemical evolution. The similar slopes of the [Mg/Fe] and [M/H] trends with age point to a similar chemical evolution at all radii.

    

\end{itemize}

\vspace{0.2cm}

Our analysis sheds new light on the disc evolution process from the perspective of the present-day solar neighbourhood population. This provides a highly precise view, but is limited by the absence of unbiased constraints at different galactic radii. \par 

In this framework, our results favour the rapid formation of an early disc settled in the inner regions (R~$<$~8.5~kpc), followed by the accretion of external metal-poor gas possibly related to a major accretion event such as the Gaia-Enceladus/Sausage one 10-12~Gyr ago \citep{helmi2018, belokurov2018, DiMatteo2019, gallart2019}. This would dilute the pre-enriched ISM abundance with a radial dependence induced by the surface density distribution of the early disc and the possible radial inflows of gas. Beyond the limiting radius of the more compact early disc, the infalling gas could have triggered the formation of stars with a chemical composition similar to the accreted Gaia-Enceladus/Sausage population, as supported by the chemical similarities between the more metal-rich Gaia-Enceladus/Sausage stars and the metal-poor thin disc population \citep{mackereth2019, feuillet2020}. \par 

In addition, our analysis supports the fact that radial migration via churning was at play in the last 10~Gyr of disc evolution \citep{georges2015a}, although the actual amplitude of the induced R$\rm _g$ changes remains largely uncertain and could be as low as 1-2~kpc. \par 

Finally, our data support a very fast chemical enrichment at early epochs, slowing down about 10~Gyr ago. The observed similar slopes in the chemical enrichment for the inner and outer regions suggest a possible equilibrium between the radial SFR dependence and the radial migration efficiency.

\begin{acknowledgements}
     We would like to thank He Zhao and Pedro Alonso Palicio for useful comments and discussions. We thank the anonymous referee for his/her constructive comments, making a considerable contribution to the improvement of the paper. The authors thank Michael Hayden for providing derived data from their 2017 paper. This work is part of the PhD thesis project within the framework of "International Grants Programme" of the Instituto de Astrof\'isica de Canarias (IAC). P.S.P. would like to thank the Centre National de Recherche Scientifique (CNRS) for the financial support. P.S.P. also acknowledge partial support from the Université Côte d'Azur (UCA). Part of this work was supported by the "Programme National de Physique Stellaire" (PNPS) of CNRS/INSU co-funded by CEA and CNES. E.F.A acknowledge financial support from the French National Research Agency (ANR) funded project “Pristine” (ANR-18-CE31-0017). A.R.B. and P.dL. acknowledge support from the ANR 14-CE33-014-01. This work has made use of data from the European Space Agency (ESA) mission {\it Gaia} (\url{https://www.cosmos.esa.int/gaia}), processed by the {\it Gaia} Data Processing and Analysis Consortium (DPAC, \url{https://www.cosmos.esa.int/web/gaia/dpac/consortium}). Funding for the DPAC has been provided by national institutions, in particular the institutions participating in the {\it Gaia} Multilateral Agreement. Most of the calculations have been performed with the high-performance computing facility SIGAMM, hosted by OCA.
\end{acknowledgements}



\bibliographystyle{aa}  
\bibliography{Santos-Peral} 

\end{document}